\documentclass[traditabstract]{aa} % for the abstract without structuration 
\usepackage{graphicx}
\usepackage{natbib}
\bibpunct{(}{)}{;}{a}{}{,}
\usepackage{epstopdf}
\usepackage{color}
\usepackage{txfonts}
\usepackage{hyperref}
\newcommand{\teff}{$T_{\rm eff}$}
\newcommand{\feh}{[Fe/H]}
\newcommand{\mh}{[M/H]}
\newcommand{\logg}{$\log g$}

\newcommand{\pmm}{$\pm$}
\newcommand{\msol}{M$_{\odot}$}
\newcommand{\rsol}{R$_{\odot}$}
\newcommand{\lsol}{L$_{\odot}$}

\newcommand{\rad}{$R$}
\newcommand{\zx}{$Z_{i}/X_{i}$}
\newcommand{\yi}{$Y_i$}
\newcommand{\mhz}{$\mu$Hz}

\newcommand{\hd}{\object{HD\,140283}}
\newcommand{\kms}{km~s$^{-1}$}
\newcommand{\fbol}{$F_{\rm bol}$}

\newcommand{\chisqr}{$\chi^2_{\rm R}$}

\newcommand{\av}{A$_{V}$}
\newcommand{\thet}{$\theta$}
\newcommand{\vmic}{$v_{\rm micro}$}
\newcommand{\vmac}{$v_{\rm macro}$}
\newcommand{\vsini}{$v\sin i$}
\newcommand{\pii}{$\pi$}
\newcommand{\hdd}{HD\,140283}
\newcommand{\lum}{$L$}
\newcommand{\corot}{CoRoT}
\newcommand{\kepler}{{\it Kepler}}
\usepackage{color}
\begin{document}

   \title{Benchmark stars for Gaia.\\ Fundamental properties of the 
 Population II star HD\,140283 from interferometric\thanks{Based on observations with the VEGA/CHARA spectrointerferometer.}, 
spectroscopic,\thanks{Based on NARVAL and HARPS data obtained within the Gaia DPAC (Data Processing and Analysis Consortium) and coordinated by the GBOG (Ground-Based Observations for Gaia) working group, and on data retrieved from the ESO-ADP database.} and photometric data\thanks{Table 12 only available in full in electronic form at the CDS via anonymous ftp to cdsarc.u-strasbg.fr (130.79.128.5) or via http://cdsweb.u-strasbg.fr/cgi-bin/qcat?J/A+A/}}

%   \subtitle{ }

   \author{O.~L.~Creevey\inst{1,2},
          F.~Th\'evenin\inst{2},
          P.~Berio\inst{2},
          U.~Heiter\inst{3},
          K.~von Braun\inst{4,5},
          D.~Mourard\inst{2},
          L.~Bigot\inst{2},
          T.~S.~Boyajian\inst{6},
          P.~Kervella\inst{7,8},
          P.~Morel\inst{2},
          B.~Pichon\inst{2},
          A.~Chiavassa\inst{2},
          N.~Nardetto\inst{2},
          K.~Perraut\inst{9},
          A.~Meilland\inst{2},
          H.~A.~Mc Alister\inst{10,11},
          T.~A.~ten Brummelaar\inst{11},
          C. Farrington\inst{11},
%          P. J. Goldfinger\inst{10},
          J. Sturmann\inst{11},
          L. Sturmann\inst{11},
          N. Turner\inst{11}
          }
\institute{Institut d'Astrophysique Spatiale, Universit\'e Paris XI, UMR 8617, CNRS, Batiment 121, 91405 Orsay Cedex, France
 \email{orlagh.creevey@ias.u-psud.fr}
%Insitut d'Astrophysique Spatiale, UMR 8XXX, Universit\'e de Paris-Sud, 91405, Orsay, France
\and
Laboratoire Lagrange, Universit\'e de Nice Sophia-Antipolis, UMR 7293, CNRS, Observatoire de la C\^ote d'Azur, 
 Nice, France
\and
Institutionen f\"or fysik och astronomi, Uppsala universitet, Box 516, 751 20 Uppsala, Sweden
\and
Max-Planck-Institute for Astronomy (MPIA), K\"{o}nigstuhl 17, 69117 Heidelberg, Germany
\and
Lowell Observatory, 1400 W. Mars Hill Road, Flagstaff, AZ, 86001, USA
\and
Department of Astronomy, Yale University, New Haven, CT 06511, USA
\and
LESIA, Observatoire de Paris, CNRS UMR 8109, UPMC, Universit\'e Paris Diderot, PSL, 5 place Jules Janssen, 92195 Meudon, France
\and
UMI-FCA, CNRS/INSU, France (UMI 3386), and Dept. de Astronom\'{\i}a, Universidad de Chile, Santiago, Chile.
\and
Universit\'e Grenoble Alpes, IPAG, F-38000 Grenoble, France;
CNRS, IPAG, F-38000 Grenoble, France
\and
Georgia State University, P.O. Box 3969, Atlanta GA 30302-3969, USA
\and
CHARA Array, Mount Wilson Observatory, 91023 Mount Wilson CA, USA}

   \date{Received ; accepted}

  \abstract
   {Metal-poor halo stars are important astrophysical laboratories that allow
us to unravel details about many aspects of astrophysics, including the
chemical conditions at the formation of our Galaxy,
understanding the processes of diffusion in stellar interiors, and 
determining precise effective temperatures and calibration of colour-effective
temperature relations.  To address any of these issues  
the  fundamental properties of the stars must first be determined.
\object{HD\,140283} is the closest and brightest metal-poor Population II halo
star (distance = 58 pc and $V=7.21$), an ideal target that allows us to 
approach these  questions, and one of a list of 34 {benchmark
stars} defined for Gaia astrophysical parameter calibration.
In the framework of characterizing these benchmark stars, 
we determined the fundamental properties of HD\,140283 (radius, mass, 
age, and effective temperature) by obtaining new interferometric and 
spectroscopic measurements and combining them with photometry from the literature.
The interferometric measurements were obtained using the visible interferometer
VEGA on the CHARA array and we determined a 1D limb-darkened angular diameter of
$\theta_{\rm 1D}$ = 0.353 \pmm\ 0.013 milliarcseconds.
Using photometry from the literature we derived
the bolometric flux in two ways: a zero reddening 
solution (\av\ = 0.0 mag) of 
\fbol\ of 3.890 \pmm\ 0.066 x 10$^{-8}$ erg s$^{-1}$ cm$^{-2}$, and 
a maximum of \av\ = 0.1 mag
solution of 4.220 \pmm\ 0.067 x 10$^{-8}$ erg s$^{-1}$ cm$^{-2}$.
The interferometric \teff\ is thus between 5534 \pmm\ 103 K and
5647 \pmm\ 105 K
and its radius is $R$ = 2.21 \pmm\ 0.08 \rsol.
Spectroscopic measurements of \hd\ were obtained using HARPS, NARVAL, and
UVES and a 1D LTE 
analysis of H$\alpha$ line wings yielded 
\teff$_{\rm spec}$ = 5626 \pmm\ 75 K.
Using fine-tuned stellar models including diffusion of 
elements we then determined %\logg\ = 3.65 \pmm\ 0.03 and 
the mass $M$ and age $t$ of \hdd.
 {Once the metallicity has been fixed, the age of the star depends on $M$, initial helium
abundance \yi\ , and  mixing-length parameter $\alpha$, only two of which are independent.
We derive simple equations to estimate one from the other two.
We need to adjust $\alpha$ to much lower values than the solar one ($\sim 2$) in order to
fit the observations, and if \av\ = 0.0 mag then $0.5 \le \alpha \le 1$. 
We give an equation to estimate $t$ from $M$, \yi\ ($\alpha$), and \av.
Establishing a reference $\alpha = 1.00$ and adopting \yi\ = 0.245 we derive a}
mass and age of \object{HD\,140283}: 
$M$ = 0.780 \pmm\ 0.010 \msol\ and $t$ = 13.7 \pmm\ 0.7 Gyr (\av\ = 0.0 mag), or
$M$ = 0.805 \pmm\ 0.010 \msol\ and $t$ = 12.2 \pmm\ 0.6 Gyr (\av\ = 0.1 mag).
Our stellar models yield an initial (interior) 
metal-hydrogen mass fraction of
$[Z/X]_i = -1.70$ 
%which would favour the reddened younger solution 
and  \logg\ = 3.65 \pmm\ 0.03.
 {Theoretical advances allowing us to impose the mixing-length parameter would
greatly improve the redundancy between $M, Y_i$, and age, while 
from an observational point of view, accurate determinations of extinction along with 
asteroseismic observations would provide critical information allowing us to overcome 
the current limitations in our results.}}

   \keywords{Stars: fundamental properties ---
     Stars: individual HD 140283 ---
     Stars: low-mass ---
     Stars: Population II ---
     Galaxy: halo ---
     Techniques: interferometry}

\authorrunning{Creevey et al.}
\titlerunning{Fundamental properties of HD 140283}

   \maketitle
%
%________________________________________________________________
%
%

\section{Introduction}

The determination of accurate and precise stellar properties 
(radius $R$, effective temperature \teff, age, \logg, ...) 
of metal-poor halo stars is a primary requirement for addressing 
many astrophysical questions from   
stellar to galactic physics.
For example, improving our knowledge of 
stellar interior processes (diffusion and Li
depletion e.g. \citealt{bonifacio98,leb00,char05,korn07,mel10})
and 
evolution models has important consequences
for determining the precise ages of stars and clusters (e.g. \citealt{gru00}),
and thus for determining the age, and formation history 
of our Galaxy, e.g. \citet{yamada13,haywood13}.

The effective temperature \teff\ is a particularly important fundamental
quantity to determine.  This has consequences for deriving masses
and ages of stars through a HR diagram analysis, and  determinations
of absolute abundances require accurate \teff\ preferably {a priori} along
with accurate knowledge of surface gravities \logg. 
However, determining \teff\
is difficult, and calibrating this scale has 
been the subject of many recent important studies, e.g. \citet{casagrande2010}.
While much progress has been made, the temperature scales still remain 
to be fully validated at the low-metallicity and high-metallicity regimes.
This is particularly important in the context of large-scale photometric 
surveys 
where many low metallicity stars are observed and 
provide important tests of stellar structure 
conditions quite different to those of our Sun, 
and hence of Galactic structure and evolution.

With recent developments from asteroseismology, 
in particular for cool FGK stars using the space-based telescopes \corot\ 
and \kepler, determinations of precise fundamental properties for large samples
of stars is possible, with direct consequences for galactic astrophysics. 
Mean densities can be determined with precisions of the
order of $\sim1$\% and in combination with other data,
masses and ages reach 
precisions of 5\% and $<10$\% \citep{metcalfe09,silva13,lebreton14,metcalfe14}.
Asteroseismic scaling relations which predict masses, radii and 
surface gravity from simple relations using global seismic quantities
\citep{bro94,kje95} have a huge potential for stellar population studies
and galacto-archaeology.  
However, while these relations have been validated in some regimes such as 
main sequence stars \citep{metcalfe14,huber12,creevey13}, 
they have yet to be validated in the metal-poor
regime (e.g. \citealt{epstein14}).  
Independent determinations of radii and masses of these stars provide
valuable tests of the widely-applicable relations.

Combining high precision distances 
that Gaia\footnote{\url{http://sci.esa.int/gaia/}} 
 will yield with robust determinations of predicted angular 
diameters from temperature-colour relations
(e.g. \citealt{kervella04b,surfb14})
yields one of the most constraining observables for stellar models: the radius.
%This also allows a validation of the {\it asteroseismic scaling relations}
%e.g. \citealt{huber12} where recently \citet{epstein14} highlighted large
%discrepancies for metal-poor stars.
These various arguments 
clearly demonstrate the need for very 
thorough studies of the most fundamental properties of nearby bright 
stars. 

The Gaia mission (e.g. \citealt{perryman05})
was successfully launched at the end of 2013.  In addition to distances 
and kinematic information, it will also deliver stellar properties for one billion objects
\citep{gaia13}.
In preparation for this mission a set of priority bright stars has been 
defined\footnote{\url{http://www.astro.uu.se/\~ulrike/GaiaSAM/}}.  
Extensive observations and analyses have been made on these benchmark stars,
with the aim of using them to define (and refine) the stellar models that will
be used for characterizing the one billion objects Gaia will observe.
One of these benchmark stars is \object{HD\,140283} and a thorough analysis of its 
fundamental properties is timely.

\hdd\ (\object{BD-10~4149}, \object{HIP~76976}, $\alpha$ = 15h 43m 03s, $\delta = -10^{\circ} 56' 01''$, $[l,b] = [356.31^{\circ}, 33.61^{\circ}]$) 
is a metal-poor Galactic Halo (Population II) star.
It is bright ($V=7.2$ mag) and nearby (58 pc) and thus a
benchmark for stellar and galactic astrophysics. 
\hd\ has been the subject of numerous studies, 
especially in more recent years where more sophisticated stellar 
atmosphere models have been used to determine accurate metallicities 
and $\alpha$ abundances (e.g. \citealt{char05}) and 
to study neutron-capture elements to understand better 
heavy element nucleosynthesis \citep{2012A&A...548A..42S,2010A&A...523A..24G,collet09}.
Concerning the fundamental properties of this star, 
\citet{bond13} published a refined parallax 
\pii\ = 17.15 +- 0.14 mas, an uncertainty one fifth of that determined by
the Hipparcos satellite (\citealt{hipparcos07}, 17.16 \pmm\ 0.68 mas), 
and assuming zero reddening
determined an age of 14.5 Gyr.
More recently, \citet{vandenberg14} determined an age for 
HD\,140283 of 14.27 Gyr.  These ages are  
slightly larger than the adopted age of the Universe (13.77
 Gyr, \citealt{bennett13}) but within their quoted error bars.
The new better precision in the parallax from \citet{bond13}
reduces the derived uncertainty
in the radius (this work) by a factor of 30\% and leaves the radius uncertainty
dominated almost entirely by its angular diameter.

The objective of this work is to determine the radius, mass, age, effective temperature, luminosity, and surface gravity of \hd.
 We present the very first interferometric measurements of this object obtained with the visible interferometer VEGA \citep{mourard11}
mounted on the CHARA array \citep{chara} 
in California, USA (Sect.~\ref{sec:observations}).  
Our observations also show the capabilities of this instrument to operate at a magnitude of V=7.2 (without adaptive optics) and at very high angular resolution ($\sim$0.35 milliarcseconds, mas), one of the smallest 
angular diameters to be measured.  We combine these interferometric data 
with multi-band photometry (Sect.~\ref{sec:fbol}), 
to derive the fundamental properties of  HD\,140283  (Sect.~\ref{sec:fundprop}). 
We then analyse high-resolution spectra (Sect.~\ref{sec:spectteff}) 
to determine its \teff\ while imposing \logg\ from our analysis.
In Sect.~\ref{sec:fparams} we use the stellar evolution code CESAM to interpret 
our observations along with literature data (Sect.~\ref{sec:extobs}) 
to infer the stellar model properties (mass, age, initial abundance).
We discuss the limitations in our observations, models, and analysis 
(Sect.~\ref{sec:discussion}) and we conclude 
by listing the next important observational and theoretical steps for overcoming these limitations.

%__________________________________________________________________
%__________________________________Observations

\section{CHARA/VEGA interferometric observations 
and the angular diameter of HD\,140283\label{sec:observations}}
 
Interferometric observations of HD 140283 were performed on four nights during 2012 and 2014
using the 
VEGA instrument on the CHARA Array and the instrument 
CLIMB \citep{sturmann10} for 3T group delay tracking. 
The telescopes E1, W1, and W2 were used for all of the observations, 
providing baselines of roughly 100 m, 221 m, and 313 m. 
Table~\ref{tab:observations} summarizes these observations.
In this table, the sequence refers to the way the observations were made where a typical 
calibrated point consists of observations of a calibrator star, the target star, 
and again a calibrator star.  We obtained a total of five calibrated points. 
The extracted $V^2$ (squared visibility) measurements for the target
star alone are the instrumental $V^2$ and these need to be calibrated by stars with
known diameters.
We used a total of three different calibrator stars whose predicted uniform disk angular diameters
in the R band are given in the caption of the table.  These were estimated using surface
brightness relations for $V$ and $V-K$ as provided by  the SearchCal tool of JMMC \citep{bonneau06}.
 {SearchCal also provides the limb-darkened-to-uniform-disk converted angular diameters, and
the latter are used for calibrating the data.}
The average seeing during the full observation sequence is given in the following column 
by the Fried parameter r0.  This parameter gives the typical length scale 
over which the atmosphere can be considered uniform.  Perfect conditions can have r0 of the order of 20 cm, while
poor conditions have r0 of approximately 5cm.  Our observations were conducted in poor to average
seeing conditions (low altitude and non-optimal observing season),  {resulting in 
larger errors on the data and the loss of some points (see below).}

\begin{table*}
\caption{\label{tab:observations} Observation log for HD\,140283 using the 
VEGA interferometer on the CHARA array in three-telescope mode 
(telescopes E1, W1, and W2).  See Sect.~\ref{sec:observations}.}
\begin{center}\begin{tabular}{lllllll}
\hline\hline
Date & Sequence  & r0 & Bands processed/ & No. of potential& Useful $V^2$ \\
dd-mm-yr & & (cm) &  width (nm) & $V^2$ points & \\
\hline\hline
2012-April-18  & C1-T-C1 & 5-8 & 3/10 & 9 & 1\\
2012-May-21 & C2-T-C1 & 7-10 & 4/10 & 12 & 12\\
2014-May-03 &  C2-T-C2-T-C2  & 8-10 & 1/20 & 6 & 4\\
2014-May-04 &  C2-T-C3-T-C1 & 7-10 & 1/20 & 6 & 2\\
\hline\hline
\end{tabular}
\end{center}
\tablefoot{The sequence contains information on the calibrator stars used and the number of calibrated points obtained where  T = target and C$N$ = Calibrator number $N$.  The column r0 contains the average Fried parameter which describes the seeing conditions.  Bands processed/width refers to the number of bands that were processed in the $\sim$45nm wavelength region and the width of each band (no overlap). The number of potential $V^2$ points depends on the number of bands processed and the number of calibrated points $\times 3$. The {\it Useful} $V^2$ are those listed in Table~\ref{tab:data}.  C1 = HD\,141378 with $\theta_{\rm UD,R} = 0.28 \pm 0.02$ mas where $\theta_{\rm UD,R}$ means the uniform disk diameter in R band, C2 = HD\,143459 with $\theta_{\rm UD,R} = 0.265 \pm 0.018$ mas, and C3 = HD\,138413 with $\theta_{\rm UD,R} = 0.301 \pm 0.022$ mas}.
\end{table*}

\subsection{Extraction of squared visibilities\label{sec:vis2}}
We used the standard V2 reduction procedure of the VEGA instrument 
as described in \citet{mourard11} to process the data.
Depending on the quality of the data, the observations were processed in 
a number of bands of different width.  
The wavelength coverage of the VEGA R/I band is $\sim 45$ nm.  
For the observations from 2012 we processed the data in 
four spectral bands of 10nm each, centred on  705nm, 715m, 725nm, and 
735nm.  
Because of the lower signal-to-noise ratio (S/N), for the 2014 observations we processed the data 
in one band of 20nm centred on
710nm\footnote{Each {\it observation} of a star (either calibrator or target) 
consists of a total of N blocks of 1000 exposures.
The calibrator observations consist of 20 blocks, while 
the target observations from 2012 consist of 60 blocks and those of 2014 consist of 30 blocks.}.
Table~\ref{tab:observations} summarizes this information under the column heading 
{\it Band processed/width}.  
The number of potential $V^2$ points depends on the number of bands processed,
the number of calibrated points, and the number of telescopes pairs (in this case three) for each observation sequence.  
In some cases (variable seeing, poor S/N on target or calibrator) the 
visibility calibration fails.
Thus, in the final column 
of Table~\ref{tab:observations} we give the number of $V^2$ points that were successfully extracted
from each observation sequence,
resulting in  a total of 19 usable $V^2$ points.
Some of the data were processed  independently and the resulting
$V^2$ varied only in the third decimal place with no consequence for the 
angular diameter derivation.
The calibrated visibilities have an associated intrinsic 
statistical error $\sigma_{\rm STAT}$ and an 
external error $\sigma_{\rm EXT}$ coming from the
uncertainty on the angular diameter of the calibrators,
 {and the total error is given 
as the quadratic sum of the two components (\citealt{mourard12}\footnote{https://www-n.oca.eu/vega/en/publications/spie2012vega.pdf}).}
By observing on different nights and by using different calibrators
we reduced the possibility of systematic errors in our analysis.
Table~\ref{tab:data} lists the squared visibilities $V^2$, 
along with the errors separated into the statistical $\sigma_{\rm STAT}$ 
and external errors $\sigma_{\rm EXT}$, 
the projected baselines\footnote{The projected baseline Bp depends on a number of 
factors including the physical distance between
the telescope pair (baseline), the angles of the baseline compared to the north-east direction
and the altitude of the object over the horizon, see e.g. Nardetto, N. PhD thesis, 2005, pg. 57 
{\tt http://tel.archives-ouvertes.fr}} Bp, and the effective wavelengths $\lambda$ of the 19 $V^2$ points.

\begin{table}
\caption{Calibrated visibilites of HD\,140283 (Sect.~\ref{sec:vis2})
\label{tab:data}}
\begin{center}
\begin{tabular}{lllllll}
\hline\hline
Date&$V^2$ & $\sigma_{\rm STAT}$ & $\sigma_{\rm EXT}$& Bp & $\lambda$ \\
(dd-mm-yy)&&&&(m) & (nm) \\
\hline%\hline

2012-April-18&0.755 & 0.222 & 0.007 & 94.087 & 715.0 \\
2012-May-21&0.790 & 0.102 & 0.006 & 93.171 & 705.0  \\
2012-May-21&0.210 & 0.097 & 0.023 & 310.487 & 705.0 \\
2012-May-21&0.659 & 0.158 & 0.038 & 221.555 & 705.0 \\
2012-May-21&0.886 & 0.133 & 0.006 & 93.171 & 715.0 \\
2012-May-21&0.213 & 0.052 & 0.023 & 310.487 & 715.0 \\
2012-May-21&0.469 & 0.079 & 0.026 & 221.555 & 715.0 \\
2012-May-21&1.043 & 0.157 & 0.007 & 93.171 & 725.0\\
2012-May-21&0.269 & 0.078 & 0.028 & 310.487 & 725.0 \\
2012-May-21&0.311 & 0.134 & 0.017 & 221.555 & 725.0 \\
2012-May-21&0.914 & 0.140 & 0.006 & 93.171 & 735.0 \\
2012-May-21&0.365 & 0.062 & 0.037 & 310.487 & 735.0 \\
2012-May-21&0.456 & 0.110 & 0.024 & 221.555 & 735.0 \\
2014-May-03&     0.541    &   0.086 &      0.042  &        221.627 &     710.0   \\
2014-May-03&     0.844    &    0.052&       0.009 &         92.743 &     710.0    \\
2014-May-03&     0.534    &    0.145&       0.033  &        218.142  &    710.0     \\
2014-May-03&     0.872   &      0.064&       0.011 &         99.990 &      710.0     \\
2014-May-04&     0.852  &      0.149    &   0.054   &       96.074 &     710.0    \\
2014-May-04&     0.245  &      0.164    &   0.047  &        312.384 &     710.0     \\

\hline\hline
\end{tabular}
\end{center}
\end{table}

\subsection{From visibilities to an angular diameter\label{sec:vis2ad}}
To determine the uniform disk and 1D angular diameters, 
we performed non-linear least-squares fits of the squared-visibility data to visibility functions
using the Levenberg-Marquardt algorithm \citep{numrec92}.
% \marginnote{is this a ratio? V2 data/visibility ? V2 data-to-visibility functions // IDL needs to be introduced here at first use}for a single star.
%The routines were implemented into IDL.  
Table~\ref{tab:angdiam} lists the uniform disk $\theta_{\rm UD}$, 
1D limb-darkened $\theta_{\rm 1D}$, and 3D limb-darkened $\theta_{\rm 3D}$ 
angular diameters and uncertainties, all explained in the following paragraphs.

\subsubsection{Uniform disk angular diameter}
Fitting the data to a uniform disk angular diameter 
ignoring the external errors 
yields $\theta_{\rm UD}$ = 0.338 $\pm$ 0.011 mas
with a $\chi^2_R$ = 0.586.
Including the external errors as 
$\sigma = \sqrt{\sigma_{\rm STAT}^2 + \sigma_{\rm EXT}^2}$ gives
$\theta_{\rm UD}$ = 0.340 $\pm$ 0.012 mas with a $\chi^2_R$ = 0.529.

\subsubsection{1D limb-darkened angular diameter \label{sec:1ddiam}}
The 1D limb-darkened disk function is given by
\begin{equation}
V = \left ( \frac{1-\mu}{2} + \frac{\mu}{3} \right )^{-1} \times 
\left[(1-\mu )\frac{J_1(x)}{x} + \mu \left(\frac{\pi}{2} \right)^{0.5}
\frac{J_{3/2}(x)}{x^{3/2}} \right] 
,\end{equation}  
where $\mu$ is the wavelength dependent limb-darkening coefficient,
$J_n$ is the Bessel function of order $n$, and 
$x = \pi B \theta_{1D} \lambda^{-1}$.
The value of $\mu$ was obtained by interpolating the linear 
limb-darkening coefficients from \citet{claret2012} (`u' in their work) for the 
stellar parameters \teff\ = 5500 and 5750 K, \logg\ = 3.75
and \mh\ = --2.10 (see Sect.~\ref{sec:fparams}).
We adopted the coefficients 
halfway between the R and I tabulated values, which correspond 
to the effective wavelengths of the measurements.
The resulting values vary between 0.4795 and 0.4870 for the range of \teff\
found in this work.
Using the values of \logg\ = 3.50 or \mh\ = --2.0 
yields limb-darkening coefficients that vary in the third decimal 
place and 
 these result in a change in \thet$_{\rm 1D}$ in the fourth decimal place.

The value of $\mu$ can in principle also be fitted, however, the 
brightness contrast is larger at the border of the projected disk 
of the star that corresponds to 
visibilities close to and just after the first zero 
point.  Our data do not extend to the first zero point nor is the 
coverage at lower visibility sufficient to 
constrain $\mu$.   Just as in the case of many other studies we 
adopt model-dependent values.

\begin{figure}
\center{\includegraphics[width = 0.5\textwidth]{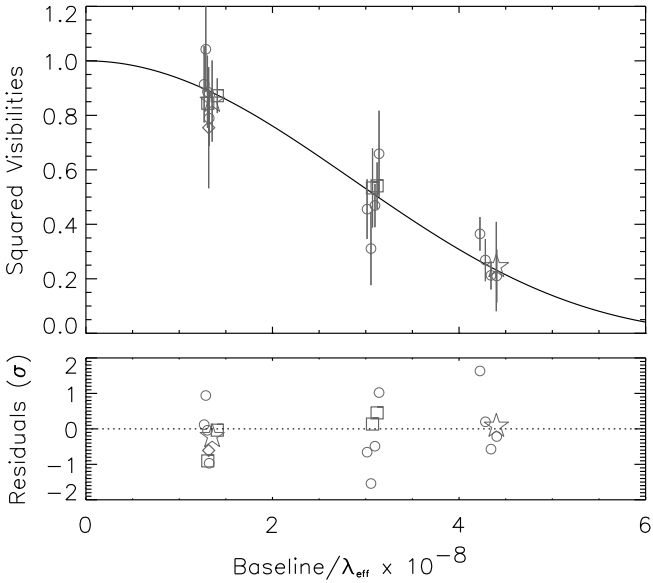}}
\caption{\label{fig:visibilityfit} {\sl Top.} 
Squared visibility measurements and the fitted visibility curve corresponding to 
a 1D diameter {$\theta_{\rm 1D} = 0.353$} mas. % \pm 0.012$ mas}.
{\sl Bottom.} Data residuals scaled by the errors on the measurements.
The symbols represent the night the data were taken:
$\diamond$ = 2012 April 18, 
$\circ$ = 2012 May 21, $\square$ = 2014 May 03,
and $\star$ = 2014 May 04 (Sect.~\ref{sec:1ddiam}).
}
\end{figure}

A fit to a 1D limb-darkened disk angular diameter
adopting the errors $\sigma = \sigma_{\rm STAT}$ yields
$\theta_{\rm 1D}$ = 0.353 $\pm$ 0.012 mas
with $\chi^2_R = 0.581$.  
The data and the corresponding visibility curve are shown in the top panel of
Fig.~\ref{fig:visibilityfit}, while the residuals scaled by the individual 
errors on the measurements are shown in the bottom panel.

%The external errors due to the externals have Gaussian errors and
%are correctly added in quadrature to the internal statistical errors
%in the data processing (the output from the data processing is in fact
%$\sigma_{\rm STAT}$, $\sigma_{\rm EXT}$ and $\sigma_{\rm TOTAL} = 
%\sqrt{\sigma_{\rm STAT}^2+\sigma_{\rm EXT}^2}$.)

 At such small angular diameters with calibrators of similar size the 
external error could have an important effect on our derived diameter.
To show explicitly the effect of this error we
performed 
Monte Carlo-type simulations.  
We simulated 10,000 sets of visibility points and using the original
statistical errors as the uncertainties in the data
we fitted these data to 
1D limb-darkened visibility functions.
A simulated set of visibility points $\mathbf{V^2} = \{ V^2_{\rm sim} \}_{i=1}^{19}$ was 
obtained by adding a random number $r$ drawn from a Gaussian 
distribution  {with a mean of zero and a standard deviation of 1,}
scaled by the external errors as follows: 
$V^2_{\rm sim} = V^2+ r\sigma_{\rm EXT}$ for each point $i$ in each set.
For the 10,000 simulated sets we obtained 10,000 fitted angular
diameters $\theta_{{\rm 1D}}$.

\begin{figure}
\includegraphics[width = 0.5\textwidth]{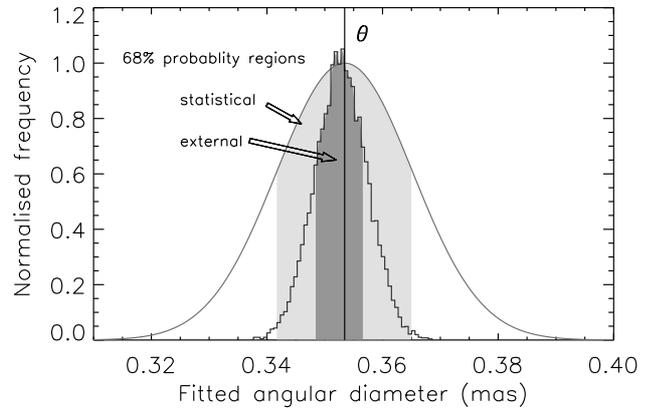}
\caption{\label{fig:simul} A Gaussian curve representing the fitted angular diameter
and the 68\% probability region ($\pm \sigma_{\theta,\rm STAT}$) due to the statistical errors (light grey shaded region).
The vertical line illustrates the centre of the Gaussian corresponding to
the fitted diameter $\theta$.
The inner distribution shows 
fitted angular diameters from 10,000 simulations by randomly adding 
the external error to the measurements.
The dark grey shaded region illustrates the 68\% probability region due to the external errors
($\pm \sigma_{\theta,\rm EXT}$, Sect.~\ref{sec:1ddiam}).}
%The corresponding error is the $\pm$ HWHM value, and the positive and negative
%contributions are shown by the horizontal line with label $\sigma_{\rm SYST}$ 
%(Sect.~\ref{sec:1ddiam}).}
\end{figure}

The distribution of angular diameters from the simulations is shown
in Figure~\ref{fig:simul} by the inner grey distribution, where
the binsize of 0.0007 mas is determined by the 
Freedman-Diaconis rule\footnote{The width of each bin $h$ is given
by $h=2q/n^{1/3}$, where $n$ is the number of points and $q$ is the 
interquartile range}. %to give64 bins.
 {The dark-shaded region corresponds to the 68\% probability region ($\pm \sigma_{\theta,\rm EXT}$,
where the subscript `$\theta$' refers to the uncertainty on the derived diameter as opposed to the 
measurement errors),
and represents a total of 0.008 mas or $\sigma_{\theta,\rm EXT} = 0.004$ (see Table~\ref{tab:angdiam}).}
We also overplot a Gaussian function representing the fitted angular diameter
$\theta_{\rm 1D}$ 
and the  {68\% probability region corresponding to the statistical uncertainty $\pm \sigma_{\theta,\rm STAT}$
(light grey shaded region).}
In this case the external errors do not contribute significantly to the total error, but
if the statistical errors were much smaller then it would be of interest to 
reduce this external error.
%The systematic error in the angular diameter is defined
%as $\pm$~half-width-half-max (HWHM) of the distribution,
%Adopting $\sigma_{\rm SYST}$ as the systematic error 
%from the simulations our angular diameter
%is given as  
%$\theta_{\rm 1D} = 0.353 \pm 0.012_{\rm STAT} {+0.006}_{\rm SYST}  - 0.004_{\rm SYST}$ mas. 
%While the external error is slightly assymetric, for the rest
%of this work we adopt a symmetric value of \pmm0.005 mas (c.f. Table~\ref{tab:angdiam}).
The uncertainty on the final adopted 1D limb-darkened diameter is obtained by 
adding the statistical and external error in 
quadrature  and this yields 
$\theta_{\rm 1D} = 0.353 \pm 0.013$ mas (see Table~\ref{tab:radius}),
in agreement with the uncertainty obtained when fitting the 
data using the total measurement errors ($\sqrt{\sigma_{\rm STAT}^2+\sigma_{\rm EXT}^2}$).
Any systematic affecting the determination of the diameter can 
only be evaluated by comparing this result with an 
independently determined angular diameter.

\begin{table}
\caption{Angular diameter determination for HD\,140283.  
All $\theta$ and $\sigma_{\theta}$ are in units of mas. 
The \chisqr\ corresponds to a fit to the statistical errors alone (Sect.~\ref{sec:vis2ad}).
\label{tab:angdiam}}
\begin{center}
\begin{tabular}{lccccccccccclllllll}
\hline\hline
&$\theta$ & $\sigma_{\theta,\rm STAT}$ & ${\sigma_{\theta,\rm EXT}}$ & $\sqrt{\sigma_{\theta,\rm STAT}^2 + \sigma_{\theta,\rm EXT}^2}$ & \chisqr\\
\hline
$\theta_{\rm UD}$ & 0.338 & 0.011 & 0.004 & 0.012 & 0.586\\
$\theta_{\rm 1D}$ & 0.353 & 0.012 & 0.004 & 0.013 & 0.581\\
{$\theta_{\rm 3D}$} & 0.352 & 0.012  & 0.004 & 0.013 & 0.582 \\
\hline\hline
\end{tabular}
\end{center}
\end{table}

\subsubsection{3D limb-darkened angular diameter}
We also calculated the limb-darkened diameter considering realistic 3D 
simulations of surface convection.   
Generally $\theta_{\rm UD} < \theta_{\rm 3D} < \theta_{\rm 1D}$ \citep{big06,big11,chiavassa09,chi12},
which is due to a smoother hydrodynamical temperature gradient in the surface
layers compared with the hydrostatic gradient,
although for non-evolved stars this difference may not be significant even
in the visible wavelengths which are more sensitive to this effect.
\citet{casa14} recently pointed out a very small discrepancy 
between their absolute \teff\ scale from
the infra-red flux method (IRFM) and measured angular diameters and attributed it  
 to the fact that 1D interferometric
diameters are usually quoted and not 3D ones.  
In most cases the precision in the measurements is
not sufficient to distinguish between the two; however, 
their work did show an almost perfect
agreement between the \teff\ derived from 
a 3D angular diameter of the metal-poor 
giant HD\,122563 from \citet{cre12b} and their IRFM value (a difference of 8 K
in \teff).
For this reason we performed a 3D analysis.
The method follows that of \citet{big11} and we refer the readers to this
paper for details.
Briefly, the 3D limb-darkened profiles are obtained by computing the full 3D monochromatic line transfer within the wavelength range of the VEGA red camera at different inclination angles. We then averaged the monochromatic limb-darkened profiles within each bandwidth (10-20nm) around the central wavelengths. The parameters that define the 3D model of granulation were 5650 K (temporal average), logg = 3.65 and [M/H] = --2.0. 
By comparing the 3D limb-darkened intensity profile directly with the 1D profile using 
a limb-darkening law and the adopted 1D coefficients,
we find an average difference of $<$1\% between the two profiles 
for over 99\% of the linear diameter of the star.
A fit to the data yields $\theta_{\rm 3D} = 0.352 \pm 0.013$ mas very close to the 1D profile.
The difference between the 3D and 1D 
is negligible for HD\,140283 since its limb darkening is very weak 
because of the lack of metals in its atmosphere.
For the rest of the work,  we adopt the 1D value to derive the stellar parameters
(see~Table~\ref{tab:radius}).

\section{Bolometric flux\label{sec:fbol}}

\begin{figure}
\includegraphics[width = 0.5\textwidth]{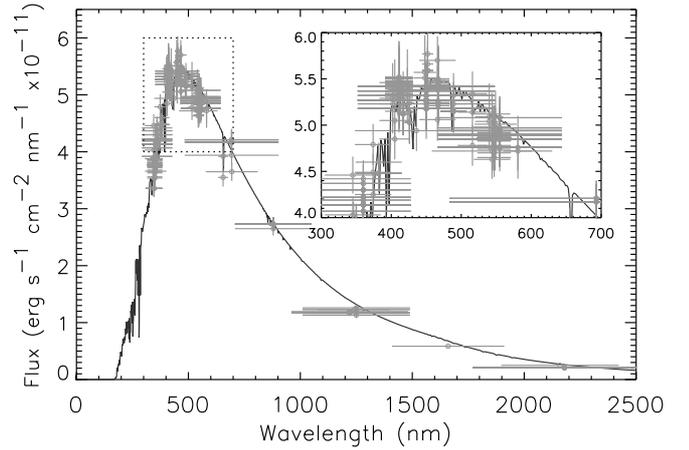}
\caption{Photometric measurements converted to absolute flux for \hd, with
an example of a fit to a model spectrum (Sect.~\ref{sec:fbol}). 
\label{fig:fbolfit}}
\end{figure}

Knowledge of the bolometric flux \fbol\ is required to calculate the 
\teff\ using the angular diameter or the luminosity $L$ using the parallax.
It can be estimated by 
using bolometric corrections with the observed $V$ magnitude of the star, 
or by using empirical formulae.
A more reliable way to determine it is by fitting  
the spectral energy distribution (SED), as long as there is sufficient 
wavelength cover.
We adopt the SED approach to determine \fbol\ and in 
this section we describe this analysis and 
compare it to literature values.

One of the biggest difficulties in determining the bolometric flux is 
the unknown reddening to the star.  
For stars that are close enough we can often make the assumption that
interstellar extinction is non-existent, or $A_V = 0.0$ mag.  
However, we also know that the distribution of gas and dust 
is not homogenous in the Galaxy
and even stars that are close may suffer some degree of reddening.
In fact, \citet{bond13} discussed this point  for \hd\ and 
remarked that by using 
\hd\ as a standard candle and assuming a small
degree of reddening yielded a distance to the globular cluster
M92 in much better agreement than without reddening. 

\hd\ is not very distant, and while 
some authors do indeed confirm 
zero reddening \citep{casagrande2010,casagrande2011,ram10}, 
  several authors have also reported 
non-zero reddening to this star.
\citet{fuhrmann98} cited several sources of extinction measurements:
$E_{b-y}$ = 0.056, 0.020, 0.043 mag from \citet{mcmillan76,sch89,hmerm90},
and $E_{B-V}$ = 0.02 mag from \citet{ryan89}.
Using  $E_{B-V} \sim 1.35 \times E_{b-y}$ and $A_V = 3.1 \times E_{B-V}$ we
obtain a range of $A_V$ between 0 and 0.24 mag.
Additionally, \citet{arenou92} estimates \av\ = 0.13 mag for \hd's galactic coordinates
and distance,
while \citet{lall14} obtain \av\ = 0.0087 mag (Puspitarini, priv. comm.). 
%, where the coefficients have been obtained from a 3D map of the Galaxy with a resolution of 20$^{\circ} \times 15^{\circ}$. 
%\citet{bjones10} also derives a non-zero extinction of $A_0$ = 0.15 - 0.75 mag
%corresponding to a 90\% confidence region, where $A_0$ is the total extinction
%(adopting a standard extinction law with R(V) = 3.1). 
%Using  $E_{B-V} \sim 1.35 \times E_{b-y}$ and $A_V = 3.1 \times E_{B-V}$ we
%obtain a range of $A_V$ between 0 and 0.24 mag, along with $A_0$ up to 
%0.75 mag.  
The adoption of one particular result for \av\ then becomes arbitrary. 
In this work we determine \av\ in the direction of the star along with \fbol\ using 
a SED fitting method, described below.  
However, we also fix \av\ = 0.0 mag and discuss our results considering 
both possible scenarios.

\subsection{Spectral energy distribution fitting methods\label{sec:fbolmethods}}

We used two different SED approaches to determine \fbol\ for 
HD\,140283, by using a compilation of 324 literature photometry
measurements of our target.   {The data were converted to flux 
measurements using the proper zero-points, e.g. \citet{fuku95}, and the flux 
measurements} are shown in Figure~\ref{fig:fbolfit}
along with a model fit\footnote{Data compilation and transformations available on request}.
The first fitting approach (Approach 1) is the method employed in \citet{vonbraun12} initially
described in \citet{vanbelle08}, and the second approach (Approach 2) is an
independent fitting method described in  Appendix A.
The two methods were developed independently of each other by
different authors.

Both approaches are based on fitting literature photometric data to libraries
of stellar spectra.  
Approach 1 was initially developed to use the Pickles library of empirical
spectra \citep{pickles98paper}, and here this is referred to as method 1A.
In this work we also replaced the Pickles library in Approach 1 (method 1A) by two other
libraries: (1B) the BASEL library of semi-empirical spectra \citep{lej97}  
and (1C) the PHOENIX library of synthetic spectra 
(see \citealt{phoenix-orig} and references therein), 
made available by \citet{phnx2013}.
Approach 2 implemented only the BASEL (2B) and the PHOENIX (2C) libraries.
The main difference between the two approaches is that the former evaluates
\chisqr\ values based on fixed points along the spectral library, 
while the latter
performs interpolation among the stellar parameters.
Below we describe the specifics of each method.
\begin{itemize}
\item[1A] 
Method 1A performs a $\chi^2$ minimization fit of the literature photometry
measurements of our target to the 
well-known empirical spectral templates
published in \citet{pickles98paper}.
The \citet{pickles98cat}  spectral templates span 0.2 - 2.5
$\mu$m and provide black-body
interpolation across wavelength ranges without data. 
Beyond 2.5 $\mu$m, extrapolation is done along a black-body curve of the
input temperature of the spectral template. 
Numerical integration of the scaled template yields the bolometric flux of the star.
The Pickles library has the advantage that it is 
purely empirical and so the spectra resemble the SED of true stars. 
However, the most metal-poor spectral template in this library has [Fe/H]~=~--0.6.
The lack of an adequate (metal-poor) 
spectral template is indicated by the relatively
high \chisqr\ values found.  It is  this high \chisqr\ value that 
motivated the implementation of the semi-empirical and pure synthetic spectral
libraries.
Table~\ref{tab:fbol_fit} lists the best \fbol, \av\ , and \chisqr\ values
corresponding to fits to the data imposing \av\ = 0.0 mag and then also
allowing \av\ to be fitted.\\

\item[1B] The BASEL library spans a longer wavelength region of 9--160,000 nm,
hence extrapolation is not necessary.
We calculated \chisqr\ values based on a set of spectral templates with
the following parameters: 
\teff\ ranging from 5250 K -- 6000 K in steps of 250 K; \logg\ = 3.0, 3.5, 4.0;
and \feh\ = --2.0, --2.5, and --3.0.
The best results are shown in Table~\ref{tab:fbol_fit} where we also indicate
the \teff\ of the best template spectrum.  In both reddening scenarios, these
correspond to \logg\ = 3.5 and
\feh\ = --2.5. %, while \teff\ ranged between 5500 and 6000 K.
We note that the non-zero reddening result corresponds to a spectral
template at the edge of our parameter space and thus it should 
be treated with caution.\\

\item[1C] Method 1C implemented the medium-resolution PHOENIX spectra, 
which cover a wavelength region of 300 --- 2500nm.  Extrapolation
beyond 2500 nm provides the missing flux, just as for method 1A.
Using the high-resolution spectra (see method 2C) 
we estimated the flux contribution between 50 and 300 nm as this
accounts for $\sim$4\% of the total flux, and thus cannot be neglected.
We inspected synthetic spectra ranging from 5400 K --- 6300 K in 
steps of 100 K, using \logg\ = 3.50, \feh\ = --2.0 and --3.0 and
$\alpha = +0.4$ ($\alpha$-enhanced elements).\\

\item[2B]  Method 2B 
implements the full BASEL library of semi-empirical
spectra and so no restrictions on the parameter ranges are imposed. 
Because of degeneracies between parameters, \logg\ and \feh\ were fixed
at 3.65 and --2.5, respectively, 
Varying \logg\ and \feh\ by 0.5 each results in insignificant changes
in the results ($\sim 0.001$ of units given, equivalent to $< 0.05\sigma$).
To determine \av\ we
fixed it at a range of input values while fitting only \teff\ and 
the scaling factor $\theta_s$ (see Appendix A) and chose the value 
that returned the best \chisqr.  Fixing \teff\ at different values
while fitting \av\ yielded the same results.
We tested that all of our results were insensitive to the initial parameters.\\

\item[2C] This final method implemented the high-resolution 
$\alpha$-enhanced (+0.4) 
PHOENIX library of spectra which span a wavelength range of 50 -- 5500 nm.
The parameter space was restricted to a \teff\ between 5000 K and 6500 K, 
[Fe/H] = --3.0 and --2.0,
and \logg\ = 3.5 -- 4.0, and in the final iteration \logg\ and \feh\ were
fixed to 3.65 and --2.5, respectively.

\end{itemize}

\subsection{Bolometric flux of HD\,140283} 
{The results from each fitting method are given in Table~\ref{tab:fbol_fit}.
For the zero reddening results, there is excellent agreement between
approaches 1 and 2, using both B and C, with differences between them  
of under 0.007 $\times 10^{-8}$ erg~s$^{-1}$ cm$^{-2}$ ($<1\sigma$).  
The disagreement between method 1A and the other methods is expected because 
of the known low metallicity of our target and the high \chisqr\ confirms this.
From these results the selection of one particular result between approaches 1 or 2 becomes arbitrary.
The difference between the results B and C are of the order of 
0.060 $\times 10^{-8}$ erg~s$^{-1}$ cm$^{-2}$ for 
both approaches 1 and 2, 
indicating simply that a systematic error arises from the 
choice of libraries of stellar spectra.
This 0.06 difference is also 
consistent with the scatter among results from the literature for a zero  reddening solution (see below).
Because we choose just one solution, we add the 0.06 difference in quadrature to the 
derived uncertainty.}

 {Considering \fbol\ with non-zero reddening, we find that 
methods 1A, 1B, and 2B are in agreement to within the errors.
While all of the \fbol\ do differ there is consistency among them.  
Method C results are systematically lower than method B, just as for the zero reddening
results, and the difference between
method 1C and 2C stem directly from the fitted \av.}

 {There is no clear evidence to have a preferred value of \av.
For this work we therefore adopt \av\ = 0.0 mag along with
one of the largest non-zero values.  We may then
consider that our results fall somewhere between the two extremes.
We choose to adopt the results from approach 2 because it 
interpolates in the template spectra to the best \teff\ for any combination
of \logg\ and \feh.
We then arbitrarily choose to use the results from method C (PHOENIX).
The adopted \fbol\ is given in the lower part of Table~\ref{tab:fbol_fit} where the extra 
source of error arising from the choice of libraries has been accounted for, 
and these values are also reproduced in the reference
Table~\ref{tab:radius} of observed stellar parameters derived in this paper.}

\begin{table}
\begin{center}\caption{Bolometric flux and interstellar reddening from fitting literature photometry-converted-to-flux measurements to 
the Pickles (A), BASEL (B), and PHOENIX (C) libraries of stellar spectra using 
two independent approaches (1 and 2).  
The final column shows the \teff\ of the best model template.
The lower part of the table summarizes the adopted \fbol\ for this work along
with an uncertainty that considers a source of error arising from
the choice of model spectra (Sect.~\ref{sec:fbolmethods}).
\label{tab:fbol_fit}}
\begin{tabular}{llccccccclllllllllllllllllllllllllll}
\hline\hline
Method &   F$_{\rm bol}$ & A$_{V}$ & $\chi^2_R$ & \teff\\
 & ($10^{-8} {\rm erg}$ ${\rm s}^{-1} {\rm cm}^{-2} $)& (mag) & & (K)\\
\hline
1A (empirical) &  3.753 \pmm\ 0.090 & ... &5.10& ...\\
1B (semi-emp)  &  3.944 \pmm\ 0.022 & ... & 1.87 & 5750\\
1C (model)     &  3.885 \pmm\ 0.011 & ... & 1.10 & 5800\\
2B (semi-emp)  &  3.951 \pmm\ 0.027 & ... & 1.35 & 5795\\
2C (model)     &  3.890 \pmm\ 0.027 & ... & 1.62 & 5742\\
\\
1A (empirical)  &    4.297 \pmm\ 0.031 & 0.170 \pmm\ 0.006 & 2.65& ...\\
1B (semi-emp)  &     4.274 \pmm\ 0.027 & 0.111 \pmm\ 0.006 & 1.68 & 6000\\
1C (model)     &     4.054 \pmm\ 0.024 & 0.056 \pmm\ 0.006 & 1.08 & 5900\\
%1C (model)     &     4.290 \pmm\ 0.027 & 0.128 \pmm\ 0.006 & 1.12 & 6000\\
2B (semi-emp) &     4.303 \pmm\ 0.032 & 0.105 \pmm\ 0.025 & 1.29 & 5644\\
%4.112 original, then small error found.
2C (model)    &     4.220 \pmm\ 0.029  & 0.099 \pmm\ 0.040 & 1.50 & 5898\\
\hline\hline
\multicolumn{3}{l}{Adopted \fbol\ (1B)}\\
%& 3.944 \pmm\ 0.064 & ...\\
%& 4.274 \pmm\ 0.066 & 0.111 \pmm\ 0.006\\
& 3.890 \pmm\ 0.066 & ...\\
& 4.220 \pmm\ 0.067 & 0.099 \pmm\ 0.040\\
\hline\hline
\end{tabular}
\end{center}
\end{table}

Several authors have also provided estimates of \fbol, either from 
direct determinations or from applications of their empirically 
derived formula.  In Table~\ref{tab:otherfbol} 
we summarize these determinations, which assume zero reddening.
Calculating a standard deviation of the results yields a scatter of 
0.053 ($10^{-8}$ erg s$^{-1}$ cm$^{-2}$).  
This value is within the representative 0.06 value that we 
add in quadrature to the uncertainty.

\begin{table}
\begin{center}
\caption{Determinations of \fbol\ by other authors assuming zero reddening.
\label{tab:otherfbol}}
\begin{tabular}{ccccccccccclllllllll}
\hline\hline
\fbol&\teff&\logg&\thet&\feh&Ref\\
($10^{-8} {\rm erg}$ ${\rm s}^{-1} {\rm cm}^{-2} $)&(K)& (cgs) & (mas) & (dex)\\
\hline
3.988 & 5755 & 3.44 & 0.330 & --2.51&1\\
3.928  & 5777 & 3.62 & 0.326 & --2.39 & 2\\
3.911 & 5842  & 3.73 & $\dots$& --2.09 &3\\
3.860 & 5691 & 4.00 &$\dots$& --2.37&4\\ 
\hline\hline
\end{tabular}
\end{center}
References. 1 \citet{gonbon09}, 2 \citet{casagrande2010},
3 \citet{casagrande2011}, 4 \citet{alonso96} 
\end{table}

\section{T$\mathbf{_{\rm eff}}$, $\mathbf{R}$, $g$, $\mathbf{L}$, and bolometric corrections\label{sec:fundprop}}
Table~\ref{tab:radius} lists $\theta_{\rm 1D}$ (hereafter \thet) and \fbol\ derived in this work.
Combining $\theta$, \fbol, and the Stephan-Boltzmann equation allows us
to solve for \teff, 
\begin{equation}
T_{\rm eff} = \left(\frac{1}{\sigma_{\rm SB}} \frac{F_{\rm bol}}{\theta^2} \right)^{0.25}
\label{eqn:teff}
,\end{equation}
where $\sigma_{\rm SB}$ is the Stephan-Boltzmann constant.
The luminosity \lum\ is calculated from 
\begin{equation}
L = 4\pi d^2 F_{\rm bol}
\label{eqn:lum}
,\end{equation}
where $d$ is the distance and we adopt the value of the parallax 
from \citet{bond13}.
The radius is calculated from 
\begin{equation}
R = \frac{\theta}{\pi},%\times ct
\end{equation}
or equivalently it is obtained from the Stephan-Boltzmann
equation $L = 4\pi  \sigma_{\rm SB} R^2 T_{\rm eff}^4$  if in this equation \teff\ is obtained from $\theta$. % is used to estimate \teff.
The surface gravity $g$ can also be estimated by imposing a value of mass $M$ 
and using the standard equation $g = GM/R^2$ where G is the gravitational
constant.  Here we adopt $M$ = 0.80 \msol\ with a conservative error of 
0.10 \msol, where the uncertainty is large
enough to safely assume no model dependence.
Table~\ref{tab:radius} provides the reference list of these  
derived stellar properties.

The bolometric magnitude is calculated directly from the luminosity 
where we adopt a solar bolometric magnitude $M_{\rm bol,\odot} = 4.74$ mag.
Considering the $V$ magnitude (see Table~\ref{tab:extobs}), our 
derived \av, along with the parallax we can then calculate the absolute $V$ magnitude $M_V$.  
The resulting bolometric corrections in the $V$ band 
are --0.179 mag and --0.168 mag for
the zero and non-zero reddening solutions, respectively.
These are all listed in Table~\ref{tab:radius}.

For the \teff\ values derived in this work we find tabulated bolometric corrections (BCs) from
\citet{flower} using the corrected coefficients from 
\citet{torres10} of --0.13 and --0.11 mag, 
respectively (zero and non-zero reddening values).
These should be compared with --0.19 and --0.18 in order to 
be consistent with the adopted bolometric magnitude of the Sun 
(Flower 
assumes M$_{\rm bol,\odot} = 4.73$, see \citet{torres10}; 
we assume M$_{\rm bol,\odot} = 4.74$).

 {Converting 2MASS $K_s$ (see Table~\ref{tab:extobs}) to Johnson-Glass $K$
using $K$ = $K_s$ + 0.032\footnote{\tiny{\url{http://www.pas.rochester.edu/~emamajek/}\\\url{memo_BB88_Johnson2MASS_VK_K.txt}}}, we derive an observed $(V-K)_0$ = 1.590 and 1.491 mag.  
The bolometric corrections tabulated for $K$ 
by \citet{houdas00} where we adopt \logg\ = 3.50 and \feh\ = --2.50 along
with our derived \teff\ are BC$_K =  1.486$ and 1.424 mag.
The bolometric correction in $V$ is then given by BC$_V$ = BC$_K$ - (V-K)$_0$
which yields --0.104 and --0.066 mag, where they adopt M$_{\rm bol,\odot} = 4.71$.
For both Flower and Houdashelt, the BCs from the zero reddening 
case are in better 
agreement with our values, although the differences are significant.

\citet{masana06} derives a bolometric correction for HD140283 of 
BC$_V = -0.202 \pm 0.053$ assuming M$_{\rm bol,\odot} = 4.74$.
This value is consistent with ours within the errors.
% to be consistent with
%their adopted M$_{\rm bol,\odot} = 4.74$.
%Our results are in good agreement with \citet{masana06}.  
%results for zero-reddening.

\begin{table*}
\caption{Stellar properties of HD\,140283 derived in this paper by adopting
$\pi$ = 17.15 $\pm$ 0.14 mas from \citet{bond13} (see Sects.~\ref{sec:observations}--\ref{sec:spectteff}).\label{tab:radius} }
\begin{center}
\begin{tabular}{lccccccccccclllllll}
\hline\hline
Property &% \multicolumn{2}{c}{Systematic error} &
Value & \\%\multicolumn{2}{c}{Systematic error} &
 & \av\ = 0.0 &   $A_V \neq 0.0$&  & \\
\hline
$\theta_{\rm 1D}$ (mas) & 0.353 \pmm\ 0.013  &\\
\fbol\ (10$^{-8}$ erg s$^{-1}$ cm$^{-2}$) & 3.890 \pmm\ 0.066 
& 4.220 \pmm\ 0.067 \\
\av\ (mag) & 0.000 \pmm\ 0.040 &  0.099 \pmm\ 0.040\\ 
$R$ (\rsol) & 2.21 $\pm$ 0.08 & \\
\logg$^{a}$ & 3.65 \pmm\ 0.06 \\
%\teff\ (K) & 5553 \pmm\ 104 &  
% 5665 \pmm\ 106 \\%& -47 & +57 \\
%$L$ (\lsol) & 4.18 \pmm\ 0.10 &4.53 \pmm\ 0.10\\
%$M_{\rm bol}$ (mag) & 3.197 \pmm\ 0.025  &3.110 \pmm\  0.024  \\
%$M_V$ (mag) & 3.382 \pmm\ 0.021 &3.270 \pmm\ 0.021\\
%BC$_V$ (mag) & --0.184 \pmm\ 0.033 & --0.160 \pmm\ 0.032\\
\teff\ (K) & 5534 \pmm\ 103 &  
 5647 \pmm\ 105 \\%& -47 & +57 \\
$L$ (\lsol) & 4.12 \pmm\ 0.10 &4.47 \pmm\ 0.10\\
$M_{\rm bol}^b$ (mag) & 3.203 \pmm\ 0.026  &3.114 \pmm\  0.024  \\
$M_V$ (mag) & 3.381 \pmm\ 0.045 &3.282 \pmm\ 0.045\\
BC$_V^b$ (mag) & --0.179 \pmm\ 0.052 & --0.168 \pmm\ 0.051\\
\hline
\teff$_{\rm spec}^c$ (K) & 5626 \pmm\ 75\\
$R_{\rm spec}^d$ (\rsol)& 2.14 \pmm\ 0.04 & 2.23 \pmm\ 0.04\\
\logg$^a_{\rm spec}$ (cm s$^{-1}$)& 3.68 \pmm\ 0.06 & 3.65 \pmm\ 0.06\\
\hline\hline
\end{tabular}
\end{center}
$^a$Imposing $M = 0.80 \pm 0.10$ \msol.
$^b$Adopting $M_{\rm bol,\odot} = 4.74$.
$^c$Adopting $\log g = 3.65 \pm 0.06$ and \feh\ = --2.46 \pmm\ 0.14.
$^d$Using $L$ and \teff$_{\rm spec}$.
\end{table*}

\section{1D LTE determination of \teff\label{sec:spectteff}}
With the primary aim of testing the agreement between the interferometric
and a spectroscopically-derived \teff\ 
for metal-poor stars, 
we performed a 1D LTE spectroscopic analysis of \hd\ using H$\alpha$ line profiles. 

\subsection{Observations and method}
Spectroscopic observations of HD\,140283 were obtained using three different
instruments (see Table~\ref{tab:obs}) 
as part of a spectral library produced by \citet{blanco14}.
The HARPS spectrograph is mounted on the ESO 3.6m telescope \citep{2003Msngr.114...20M}, and the spectra were reduced by the HARPS Data Reduction Software (version 3.1).
The NARVAL spectrograph is located at the 2m Telescope Bernard Lyot \citep[Pic du Midi,][]{NARVAL}. The data from NARVAL were reduced with the Libre-ESpRIT pipeline \citep{donati97}.
The UVES spectrograph is hosted by unit telescope 2 of ESO's VLT \citep{2000SPIE.4008..534D}. Two UVES spectra were retrieved for HD~140283, one from the Advanced Data Products collection of the ESO Science Archive Facility (reduced by the standard UVES pipeline version 3.2, \citealt{2000Msngr.101...31B}), and one from the UVES Paranal Observatory Project (UVES POP) library \citep[processed with data reduction tools specifically developed for that project]{2003Msngr.114...10B}.
All spectra have been convolved to a spectral resolution of $R=\lambda/\Delta\lambda=70000$, from a higher original resolution. 

\begin{table}[h]
   \caption{Information on observed spectra used for the fit: instrument/archive name, date of observation, mean signal-to-noise ratio (S/N). For instrument/archive references see text.}
\label{tab:obs}
\centering
\begin{tabular}{lll}
\hline\hline
Instrument & Date & Mean S/N \\
%\hline
%\multicolumn{3}{c}{HD~140283} \\
\hline
HARPS           & 2008-Feb-24, 2008-03-06  & 535 \\
NARVAL          & 2012-Jan-09              & 250 \\
UVES POP        & 2001-Jul-08              & 835 \\
UVES Archive    & 2001-Jul-09              & 290 \\
\hline\hline
\end{tabular}
\end{table}

The spectrum calculations and the fitting procedure were done with the Spectroscopy Made Easy (SME) package\footnote{\url{http://www.stsci.edu/~valenti/sme.html}} \citep[Version 360, 2013 June 04,][]{Vale:96,2005ApJS..159..141V}.
This tool performs an automatic parameter optimization using a Levenberg-Marquardt chi-square minimization algorithm.
Synthetic spectra are computed by a built-in spectrum synthesis code for a set of global model parameters and spectral line data. A subset of the global parameters is varied to find the parameter set which gives the best agreement between observations and calculations.
The required model atmospheres are interpolated in a grid of MARCS models \citep{2008A&A...486..951G}.

An important step in spectrum analysis with SME is the definition of a line mask, which specifies the pixels in the observed spectrum that should be used to calculate the chi-square.
We defined three different line masks, covering different windows in the wings of H$\alpha$ which are free of telluric and stellar lines (judged from the observed spectra).
Line mask~1 is similar to that used by \citet{2011A&A...531A..83C} -- four narrow windows on either side of the line centre, reaching rather far out into the wings.
Line mask~2 is similar to that used for the SME analysis of the UVES spectra obtained in the Gaia-ESO survey (Bergemann et al., in preparation) -- two somewhat broader windows on either side, closer to the line centre than for Line mask~1.
Line mask~3 is similar to that used by \citet{ruchti13}, covering as much of the line wings as possible between 1 and 10~\AA\ from the line centre on both sides.
The line masks are visualized in Fig.~\ref{fig:obs}, together with the NARVAL spectrum.

In an independent step from the parameter optimization, SME normalizes the spectrum to the local continuum by fitting a straight line through selected points in the observed spectrum defined by a continuum mask.
We defined two sets of continuum masks. The narrow continuum mask has two windows about 22 and 29~\AA\ from the line centre on the blue and red side, respectively. The wide continuum mask has two windows about 32 and 34~\AA\ from the line centre on the blue and red side, respectively. The continuum mask windows are between 0.5 and 0.9~\AA\ wide.

\begin{figure*}
   \begin{center}
\includegraphics[width = 0.9\textwidth]{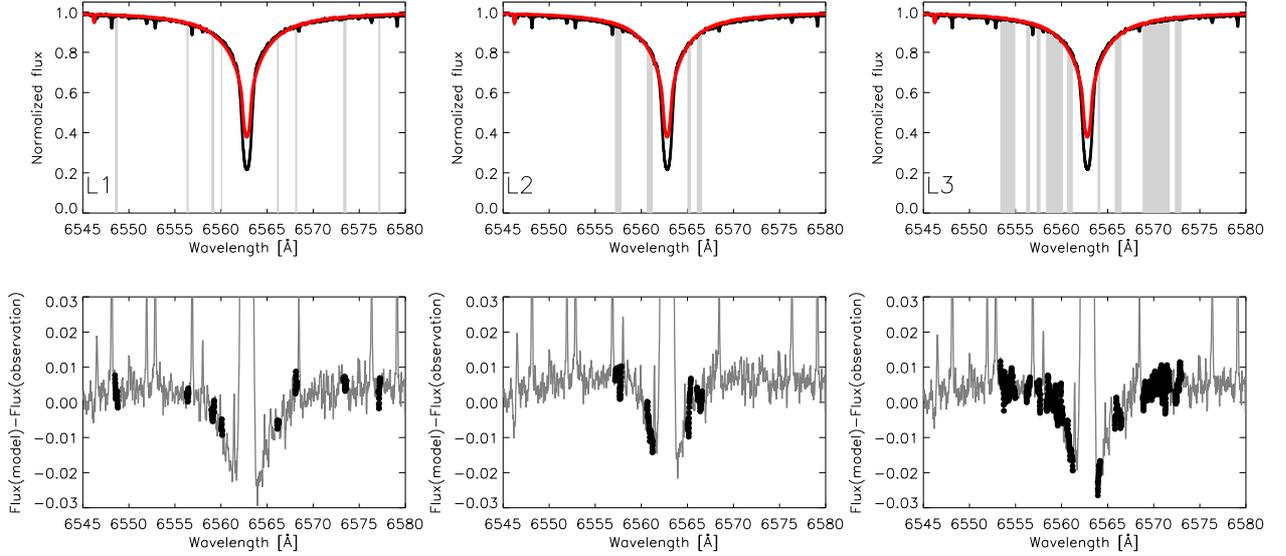}
   \end{center}
   \caption{{\sl Top: } 
Observed NARVAL spectrum (black) of the H-$\alpha$ profile of 
\hd\ and fitted spectrum (red) using the three narrow continuum masks (shaded regions) for \logg\ = 3.65 and [Fe/H]~=~--2.46.  
{\it Bottom:} Residuals between the observed and fitted spectrum for 
each line mask, with the black dots indicating the line masks. See Sect.~\ref{sec:spectteff} for details.}
\label{fig:obs}
\end{figure*}

\begin{table}[h]
   \caption{Atomic data for spectral lines contributing to flux in Line mask~3 for HD~140283}
\label{tab:lines}
\begin{center}
\begin{tabular}{lrrrrl}
\hline\hline
Species & Wavelength [\AA] & $E_{\rm low}$ [eV] & $\log gf$ & Reference \\ % & Depth & Flag
\hline
Ti II & 6559.588 & 2.048 & -2.175 & 1 \\                  % & 0.28 & N
Fe I  & 6569.214 & 4.733 & -0.380 & 2 \\  % & 0.10 & U
Ca I  & 6572.779 & 0.000 & -4.240 & 3 \\  % & 0.18 & Y
%Ti I  & 6596.534 & 3.583 & 0.100 & 1 \\ % & 0.20 & non-clean
\hline\hline
\end{tabular}
\end{center}
\tablefoot{$E_{\rm low}$ is the excitation energy of the lower level, $\log gf$ the oscillator strength. The last column gives the reference for the $gf$-value.  The oscillator strength $f$ is a measure of the strength of an atomic transition between two excited levels caused by interaction with photons. The line absorption coefficient is directly proportional to $f$, and $g$ is the statistical weight of the lower level determined by its rotational quantum number. For further details see e.g. \citet{2010EAS....43...91W} or \citet{2005oasp.book.....G}.  References. 1 = \citet{K10}, 2 = \citet{1974AaAS...18..405M}, 3 = \citet{1997ZPhyD..41..125D}.}
\end{table}

% line list
The atomic data were taken from the line list compiled for the Gaia-ESO public spectroscopic survey \citep{2012Msngr.147...25G}. In the spectral interval from 6531 \AA\ to 6598 \AA, the list contains 70 atomic lines in addition to H$\alpha$, but only three of them have a possible contribution to the flux in the Line mask~3 wavelength regions for HD~140283. The atomic data for these lines are given in Table~\ref{tab:lines}.
The $gf$-value used for the H$\alpha$ line is 0.710 \citep[see \citealt{2009JPCRD..38..565W}]{2008Baker}.
The broadening of the H$\alpha$ line by collisions with neutral hydrogen was calculated using the theory of \citet{2000A&A...355L...5B}, which is an improvement on the theory of \citet{1966PhRv..144..366A}.
The more recent calculations by \citet{2008A&A...480..581A} extend the description of the self-broadening for H$\alpha$ even further, but the resulting line profile is very similar to that using the \citet{2000A&A...355L...5B} theory. The differences are smaller than the uncertainties in the observations.

\subsection{Results}
 For the present study we adopted the observed \logg\ derived in this work (\logg\ = 3.65 \pmm\ 0.06, see Sect.~\ref{sec:fundprop}).  For \feh\ we adopted the mean and standard deviation of the reported
values from the literature between 1990 and 2011, as listed in the PASTEL catalogue \citep{soubiran10}, of \feh\ = --2.46 \pmm\ 0.14.  Line broadening from microturbulence \vmic, macroturbulence, \vmac, and \vsini\ (rotation) were all fixed, \vmic\ = 1.3 \kms, \vmac\ = 1.5 \kms, and
\vsini\ = 5.0 \kms, and changing them had no impact on the final result.

 {The initial value for the free parameter \teff\ was chosen to be 6000~K (using an initial value of 5000~K resulted in the same solution for \teff, to within 10~K).
The results using \logg\ = 3.65 and \feh\ = --2.46, rounded to the nearest 10 K, 
are given in Table~\ref{tab:HD140283fit} for each mask and each observation set along with its
corresponding \chisqr. The mean and scatter among these values is 
\teff\ = 5642 \pmm\ 63 K.  }

% Table
%
\begin{table*}
   \caption{Best-fitting \teff\ values rounded to 10~K and reduced chi-square ($\chi^2_R$) for the different observations and masks for HD~140283. The first line specifies the instrument used for the observed spectrum. The column $L$ specifies the line mask (see text and Fig.~\ref{fig:obs}).  The final row gives the 
scatter $s$ arising from the different masks.}
\label{tab:HD140283fit}
\centering
\begin{tabular}{lcccccccccccrrrrrrrr}
\hline\hline
%% table header
 & \multicolumn{2}{l}{HARPS} & \multicolumn{2}{l}{NARVAL} & \multicolumn{2}{l}{UVES POP} & \multicolumn{2}{l}{UVES Archive} \\
$L$ & \teff~(K) & $\chi^2_R$ & \teff~(K) & $\chi^2_R$ & \teff~(K) & $\chi^2_R$ & \teff~(K) & $\chi^2_R$ \\
\hline
%% table data
\multicolumn{9}{l}{narrow continuum mask} \\
\hline
1 & 5640 & 10.4 & 5750 & 1.6 & 5670 & 11.2 & 5610 & 1.2\\ %
2 & 5540 & 12.7 & 5640 & 3.4 & 5600 & 33.6 & 5560 & 3.7\\ %
3 & 5600 & 14.3 & 5710 & 3.5 & 5660 & 41.9 & 5620 & 4.7\\ %
\hline
\multicolumn{9}{l}{wide continuum mask} \\
\hline
1 & 5630 & 9.4  & 5780 & 2.3 & 5700 & 19.5 & 5630 & 1.8\\ %5680
2 & 5540 & 12.6 & 5670 & 3.7 & 5630 & 36.2 & 5580 & 3.8\\ %5590  :1+2:5635        ;;; tt_1-2=5624.5
3 & 5590 & 13.3 & 5740 & 4.2 & 5700 & 54.3 & 5634 & 5.7\\ %5657.5   ;total=5642.5 ;;; tt_1-3 = 5632.92
 \hline
$s$ (K) &  44 & &50 & & 37 & &  28\\
\hline\hline
\end{tabular}
\end{table*}

 {As can be seen from Table~\ref{tab:HD140283fit}, the best and comparable \chisqr\ values 
are obtained using the NARVAL and UVES archive data, although the differences between
their \teff\ are of the order of 100 K.  The observations that show the least sensitivity to 
the different masks are also the UVES archive data with a scatter of only 28 K, 
while those with the most sensitiviy are the NARVAL data (50 K).  }

 {To illustrate the effect of changing \logg\ and \feh, in Table~\ref{tab:spectlogfeh} we 
show the mean \teff\
of the 24 results (averaging over the four sets of observations and six masks each) 
for nine combinations of 
\logg\ and \feh\ considering the adopted errors in these parameters.
The final column shows the typical scatter $s$ found for each value given in that row. 
%, where 
%very similar values were found for different values of \logg\ (to within 3K) but not for \feh.  
The \teff\ are almost identical for \feh~=~--2.46 and --2.32, while the results for \feh~=~--2.60
decrease by $\sim30$ -- 40 K.  
%The mean and standard deviation of the 9$\times24$ \teff\ are 5628 \pmm\ 73 K.
To get a more realistic estimate of the error due to using different values of \logg\ and \feh, we
performed Monte Carlo simulations and determined 24 values of \teff\ (four observations, six masks)
by adopting a randomly chosen \logg\ and \feh, where these were drawn from a normal 
distribution shifted and scaled by the means and standard deviations as given above.
%$\sigma (\log g) = 0.06$ and $\sigma ([{\rm Fe/H}]) = 0.14$ and their means 
%as argued above.
The simulations were repeated 100 times.
The mean and standard deviation of the 2400 fitted \teff\ is 
5626 \pmm\ 75 K\footnote{Table available on request.}. % from the 2400 determinations.
This is the value that we adopt, and it is listed in the reference Table~\ref{tab:radius}.}
%We note that eliminating the NARVAL data this value reduces to 5600 \pmm\ 59 K

\begin{table}
\begin{center}\caption{\label{tab:spectlogfeh} Mean \teff\ values fixing \logg\ and \feh\ at their
mean and $\pm 1\sigma$ values.
The final column shows the typical scatter $s$ found for each of the three results given in that row.
All results are given in K.}
\begin{tabular}{lcccccccccccccccccc}
\hline\hline
$\log g$ & 3.59 & 3.65 & 3.71 & $s$ \\
$[$Fe/H$]$\\
\hline
--2.60&   5627 & 5601 & 5586 & 84\\
--2.46 &  5659 &  5642 & 5622 & 64\\ 
--2.32 & 5658 & 5640 & 5621 & 63\\ 
\hline\hline
\end{tabular}
\end{center}
\end{table}

 {The interferometric \teff\ falls between a mean value of 5534 K and 5647 K \pmm\ 
105 K depending on the amount of reddening between the top of our Earth's atmosphere
and the star.  
We can then conclude that the 1D LTE determination
of \teff\ from H$\alpha$ line wing fitting yields results that are consistent 
with the interferometric values and are thus valid for stars at this lower
metallicity regime within the typical uncertainties for spectroscopic observations.}

Adopting our spectroscopic \teff, along with \fbol, and \pii, we 
determine $R$ using the Stefan-Boltzmann equation,
$R$ = 2.14 \pmm\ 0.04 \rsol\ and 2.23 \pmm\ 0.04 \rsol, 
without and with reddening, respectively, or the angular diameter
of 0.341 mas and 0.356 mas, respectively, ignoring the distance.
%There is excellent agreement with our interferometrically derived angular diameter
%(0.353 mas) and thus radius (2.21 \rsol), for the 
%reddened solution. % and within 1$\sigma$ for the non-reddened solution.
Again assuming $M = 0.80 \pm 0.10$ \msol, 
we obtain an evolution-model-independent determination of \logg,
3.68 \pmm\ 0.06 and 3.64 \pmm\ 0.06, respectively, where the error is 
dominated by the imposed error on mass. 
These spectroscopically derived values are also given in Table~\ref{tab:radius}.

\section{Previously observed stellar parameters\label{sec:extobs}}

\begin{table}
\caption{External measurements for \hd.\label{tab:extobs}}
\begin{center}
\begin{tabular}{lllllllll}
\hline\hline
$V$ (mag) &   7.21 $\pm$ 0.01 \\
$K_s$ (mag) & 5.588 \pmm\ 0.017  \\
$\theta_{\rm 1D, pred}$ (mas) & 0.325/0.321$^\dagger$\\
$\pi$ (mas) & 17.15 $\pm$ 0.14\\
\feh\ (dex, LTE) & --2.46 \pmm\ 0.14\\
\feh\ (dex, NLTE) & --2.39 \pmm\ 0.14\\
$[\alpha /{\rm Fe} ]$ (dex) & +0.40 \\
\mh$^\star$ (dex) & --2.10 $\pm$ 0.20\\
 $^6$Li/$^7$Li & 0.018 \\
A(Li) (dex) & 2.15 --- 2.28 \\ \hline\hline
\end{tabular}\end{center}
$^\dagger$Predicted angular diameter assuming \av\ = 0.0 and \av\ = 0.1 mag.
$^\star$Adopting the coefficients of \citet{gn93} in line with the abundances 
used in the modelling section.
\end{table}

Using the General Catalogue of Photometric Data \citep{gcpd95}, 
we found several sources of $V$ magnitudes
for \object{HD\,140283} more recent than 1980. 
These are 7.194, 7.212, 7.21, 7.22, 7.22, and 7.20 mag
\citep{grier80,cousins84, norris85,1986EgUBV........0M,carlat87, upgren92}.
Adopting the mean and standard deviation of these we obtain
$V = 7.21 \pm 0.01$ mag. % as the $V$ magnitude
Using all of the available photometric measurements that date back as 
far as 1955 we obtain a mean $V = 7.214 \pm 0.018$ mag with a mean error
of 0.051 mag (these were the data used in  Sect.~\ref{sec:fbol}).
The infrared magnitudes from the 2MASS catalogue \citep{2mass06} are
$H = 5.696 \pm 0.036$ mag and $K_s = 5.588 \pm 0.017$ mag. 
The $V$ and $K_s$ magnitudes are given in Table~\ref{tab:extobs}.

We can apply the surface-brightness-colour relations calibrated by 
\citet{kervella04b} and \citet{surfb14} to HD\,140283, 
to check the validity of these relations in the low metallicity regime. 
Combining the $B$ magnitude from \citet{morel78} $B = 7.71$
with the $V$ and the infrared magntiudes $H$ and $K_s$, 
the \citet{kervella04b} relations yield 
consistent 1D limb-darkened angular diameters between the different colour indices 
of 
$\theta_{\rm 1D}=0.321$ mas, % \pm 0.010$\,mas, 
while considering a reddening correction corresponding to $A_V = 0.10$ mag.
Considering no reddening correction at all results in a marginal change of the 
predicted angular diameter ($\theta_{\rm 1D}=0.325$ mas). % \pm 0.010$\,mas).
Using the more newly calibrated relations from \citet{surfb14} we also obtain 
$\theta_{\rm 1D} = 0.325$ mas, where such relations yield a scatter of the order of 
5\%, which in this case corresponds to an uncertainty of approximately 0.016 mas.  
Our measured value is $\theta_{\rm 1D}=0.353 \pm 0.013$, a difference of 
just over 2$\sigma$.
This difference could perhaps be explained by the surface brightness relations
being calibrated with more metal-rich stars.

\citet{jofre14} recently presented a homogenous analysis of 34 FGK benchmark stars
to derive their metallicities. 
They use several methods to determine \feh\ 
including the SME method presented here.  
For \hd\ they adopt \teff\ = 5720 K and \logg\ = 3.67 
and then derive an NLTE corrected \feh\ by combining individual line
abundances of neutral lines.  %and subsequently consider NLTE corrections.
They also study the sensitivity of \feh\ to \teff\ and find that a difference of 120 K
results in a \feh\ difference of 0.04 dex.
Applying this correction to the spectroscopic \teff\ found here ($\sim$--90 K) yields 
a NLTE \feh\ = --2.39 and an LTE \feh\ = --2.46, in line with the value used in Sect.~\ref{sec:spectteff}, but with lower uncertainties of 0.07 dex.
We note that in the initial analysis presented in their paper, individual 
methods based on a global fit to the spectrum yielded 
slightly lower LTE \feh\ between --2.44 and --2.57.
These last results have also been found by \citet{2010A&A...523A..24G} 
and \citet{2012A&A...548A..42S}: \feh\ = --2.59 and --2.60 
adopting \teff\ = 5750 K and \logg\ = 3.70.

\citet{the98cat} derive \feh\ and abundances for the $\alpha$ elements using a
\teff\ = 5600 K and \logg\ = 3.2 dex; % (microturbulence = 1.4 \kms);
\feh\ = --2.50, [Mg/H] = --2.05, [Si/H] = --2.10, [Ca/H] = --2.10, and
[Ti/H] = --2.35, 
giving an average/median value of [$\alpha$/Fe] = +0.35/+0.40.
Adopting the NLTE \feh\ of --2.39 (by adapting the \citealt{jofre14} value)
along with [$\alpha$/Fe] from \citet{the98cat} yields a global
metallicity \mh\ = --2.12/--2.08 adopting the coefficients 
provided by \citet{salaris93} or \mh\ = --2.14/--2.10 adopting
those provided by \citet{gn93} as given in \citet{yi01}.
To account for the uncertainty on the $\alpha$ elements 
along with the 0.14 dex on \feh\ as discussed in Sect.~\ref{sec:spectteff}, 
we adopt a conservative 0.20 dex as an error bar on \mh.
Finally, 
\citet{bonifacio98} derive a Li abundance of A(Li) = 2.146, and %1.4 $\times 10^{-11}$. 
\citet{char05} derive A(Li) = 2.26 -- 2.28.
These parameters are all summarized in Table~\ref{tab:extobs}.

%Their Fig.~12 depicting A(Li) versus \teff\ for a sample of about 100 metal-poor halo stars shows that for \teff\ lower than $\sim$5500 K, A(Li) $< 1.5$ in general.  A value of 5665 K which we derive by assuming non-zero reddening would be more compatible with the higher observed A(Li).

%__________________________________________________________________

\section{Interpretation of new data using stellar evolutionary models \label{sec:fparams}}

%__________________________________________________________________

We used the 
CESAM stellar evolution and structure code (CESAM2k) 
\citep{mor97,ml08} to derive the 
model-dependent
properties of mass $M$, age, $t$, initial helium abundance mass fraction $Y_i$, and
initial chemical composition $Z_i/X_i$, where $Z$ and $X$ denote
heavy metals and hydrogen with $Z_i+Y_i+X_i=1$.  
The initial abundances are at zero-age main sequence (ZAMS).

\subsection{CESAM2K physics\label{sec:models}}
The input physics of the models consist of
the equation of state by \citet{eff} with Coulomb corrections, 
the OPAL opacities \citep{ri92} 
supplemented with \citet{af94} molecular opacities.
The p-p chain and CNO-cycle nuclear reactions were calculated
using the NACRE rates \citep{ang99}.
We adopted the solar abundances of \citet{gn93} 
($Z/X_{\odot} = 0.0243$) 
and used the MARCS model atmospheres \citep{marcs03} for a metallicity of 
[M/H] = --2.0  {for the boundary conditions and reconstruction of the atmosphere}.
Convection in the outer envelope 
is treated by using the  mixing-length theory described by \citet{egg72},
where $l = \alpha H_p$ is the mixing-length that tends to 0 as the 
radiative or convective borders are reached, $H_p$ is the pressure scale
height,  and 
$\alpha$ is an adjustable parameter.
Calibrating this value for the solar parameters we obtain 
$\alpha \sim 2.0$. % \red{checking solar value}. 
Microscopic diffusion was taken into account 
and follows the treatment described
by \citet{bur69}, and 
extra mixing is included by employing 
a parameter, $Re_\nu$, as prescribed
by \citet{morelthev02} which slows down the depletion of helium and 
heavy elements during evolution. 
\citet{nord12} showed that
the abundance difference between turnoff stars and red giants
in the globular cluster NGC~6397 (\feh\ = --2.0)
is typically 0.2 dex.
We chose a value of the extra-mixing parameter $Re_\nu$ = 5, to 
produce a similar abundance difference.
Figure~\ref{fig:diff} illustrates the evolution from ZAMS of the surface metallicity
with \teff\ by employing $Re_\nu$ = 5.
As can be seen the initial surface metallicity is quite different to the observed
one, which is illustrated by the arrows.
For the metallicity of the star we adopt a [M/H] ($\equiv [Z/X]_s$) = --2.10 \pmm\ 0.20 as 
argued in Sect.~\ref{sec:extobs}.  
Using $Z/X_{\odot}$ = 0.0243, without considering diffusion of 
chemical elements in the models, 
we obtain $Z_i/X_i$ = 0.00022 + 0.00013 -- 0.00008. 
%Adopting less conservative errors bars on the metallicity of 
%0.1 dex reduces these uncertainties to \pmm\ 0.0005.
From Fig.~\ref{fig:diff} it can be seen that the observed metallicity is 
lower than the initial one, and so the first constraint we have
on the initial metallicity is $Z_i/X_i  > 0.00022$.

\begin{figure}
\includegraphics[width=0.45\textwidth]{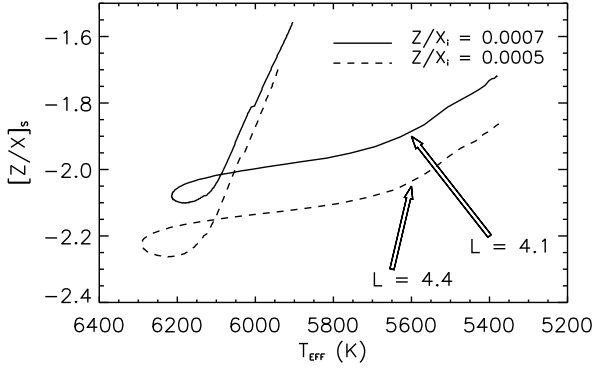}
\caption{\label{fig:diff} The evolution of the observed (surface) metal abundance,
shown as a function of \teff, for a 0.80 \msol\ model with $Y_i$ = 0.245 
including the effects of diffusion and extra mixing for 
two initial metal-hydrogen ratio mass fractions ($Z_i/X_i$ = 0.0005, 0.0007).
At \teff\ = 5600 K, we highlight the values of the luminosity for both of the
models. See Sect.~\ref{sec:models} for details.}
\end{figure}

\begin{figure*}
\center{\includegraphics[width=0.48\textwidth]{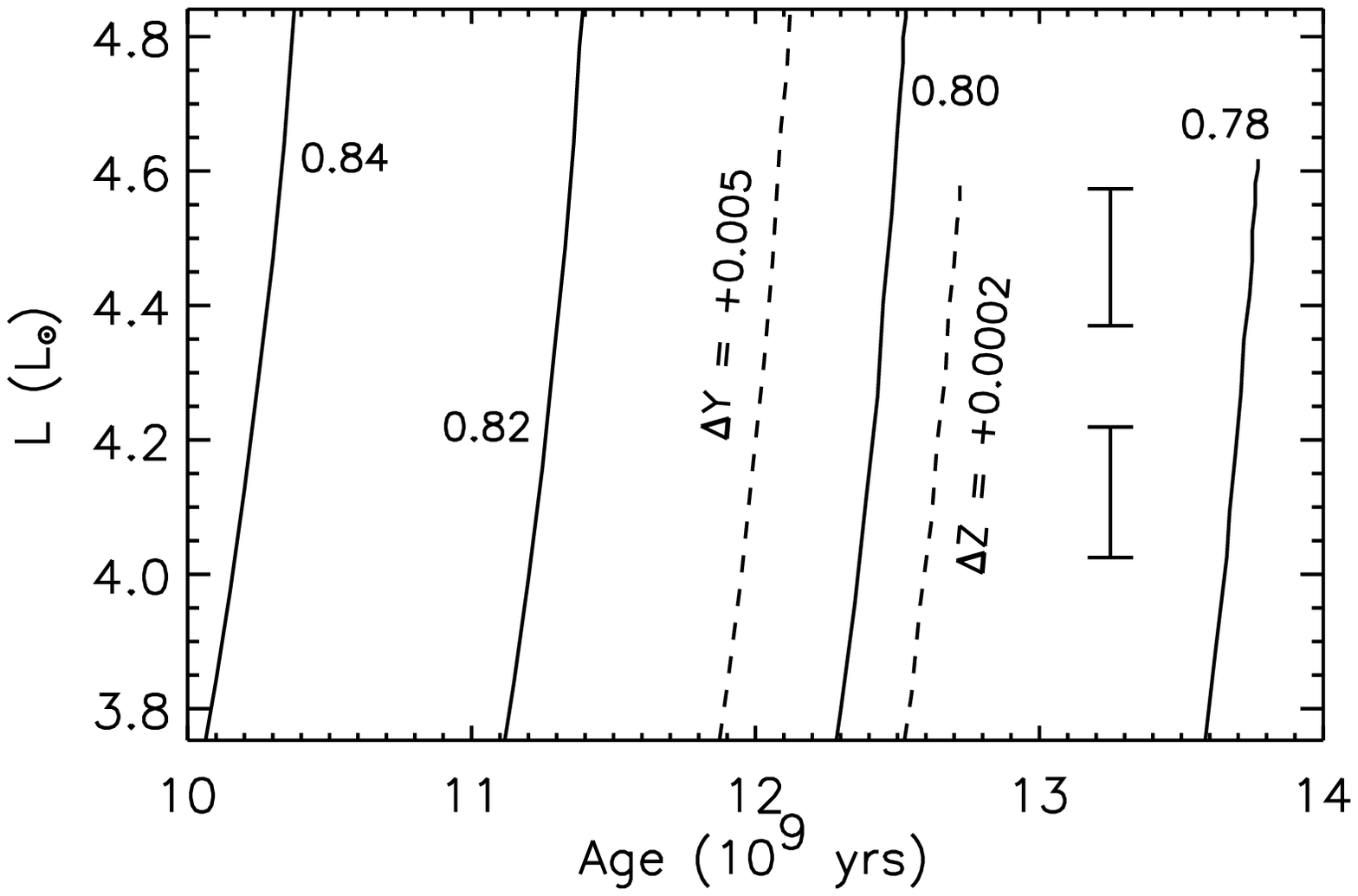}
\includegraphics[width=0.48\textwidth]{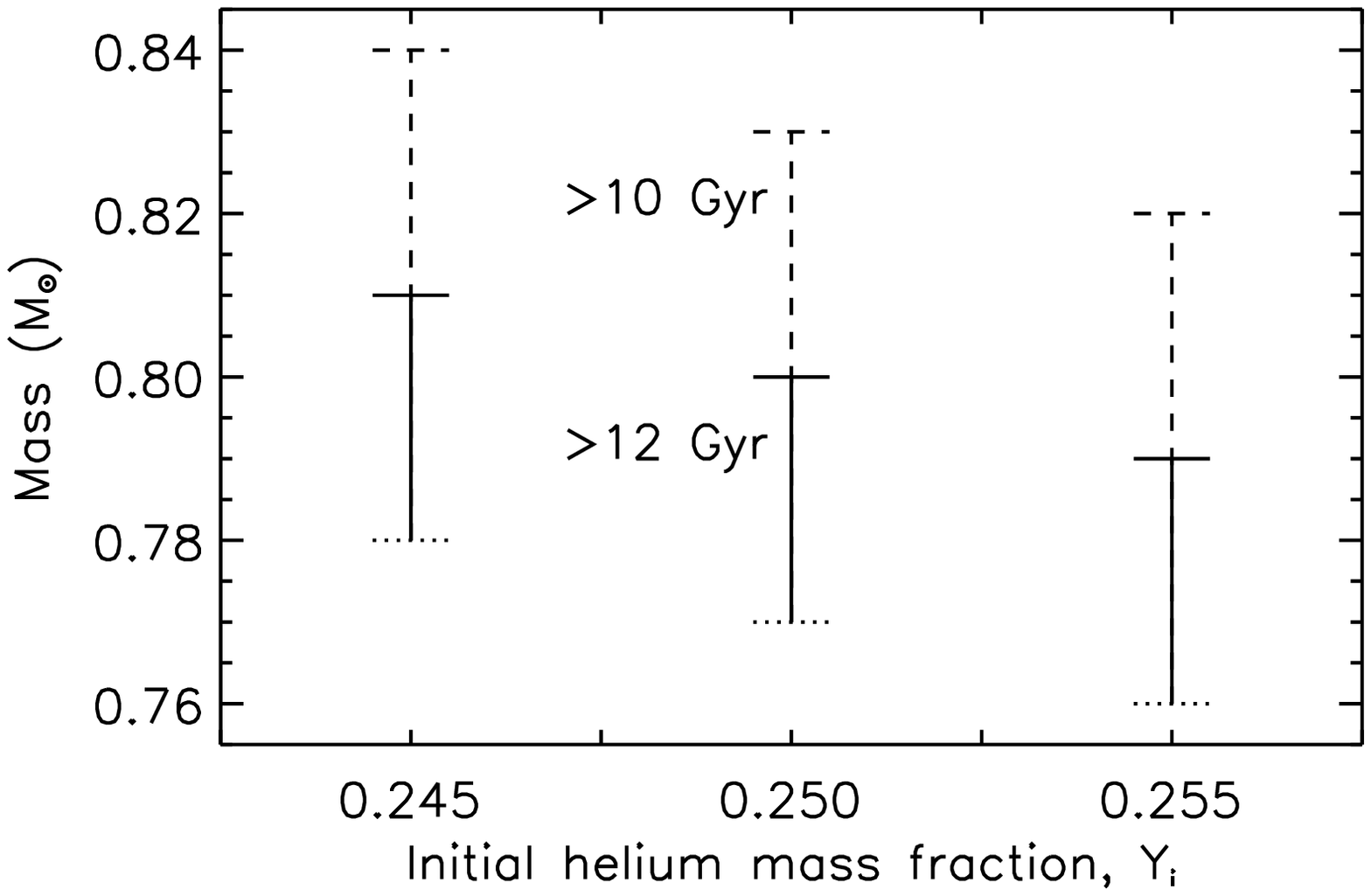}}
\caption{{\sl Left:} Evolutionary tracks for models of different masses (labelled by
their solar masses) for $Y_i$ = 0.245 and $Z_i/X_i$ = 0.0005 plotted as a function
of age.  The dashed lines illustrate changing the $M =0.80$ \msol\ 
model in $Y_i$ ($\Delta Y$) and  $Z_i/X_i$
($\Delta Z$).  The unreddened and reddened luminosity constraints are
indicated by the error bars.
{\sl Right:} Approximate correlations between $M$ and the initial helium abundance $Y_i$
assuming an age  $>12$ Gyr (continuous) and then extending it to $>10$ Gyr (dashed),
given the luminosity constraint. See Sect.~\ref{sec:amlum} for details.
\label{fig:amlum}}
\end{figure*}

Each stellar evolution model is defined by a set of input model parameters ---
mass $M$, initial helium content $Y_{i}$, 
initial metal to hydrogen ratio \zx,
age $t$, the mixing-length parameter $\alpha$, and the extra-mixing
coefficient $Re_{\nu}$ ---
and these result in model observables, such as a model \teff\ and a model
$L$.
By varying the input parameters we 
aimed to find models that fitted the derived $L$, \teff\ , and \mh\ as given in 
Tables~\ref{tab:radius} and \ref{tab:extobs}.

\subsection{Approximate parameters of HD\,140283 using $L$\label{sec:amlum}}

\begin{figure}
\center{\includegraphics[width = 0.45\textwidth]{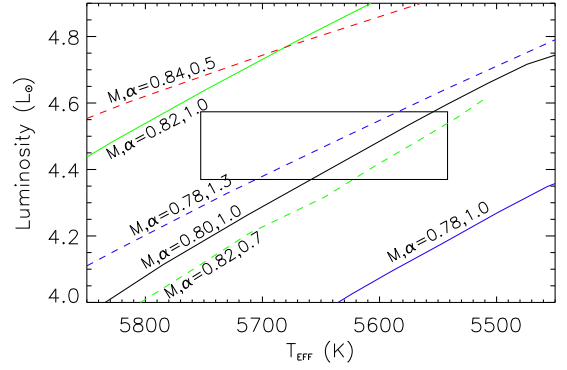}}
\caption{\label{fig:alpha}Correlations between $M$ and $\alpha$ for the 
reddened solution (error box) imposing \yi\ = 0.245, illustrating
the plausible ranges for these parameters imposed by stellar evolution.  
Similar colours have the same mass and the continuous lines have $\alpha = 1.0$.
See Sect.~\ref{sec:alpha} for details.}
\end{figure}

The luminosity $L$ 
provides a very important constraint on the evolution stage of the star.
Because it is a measure of the energy production in the star, we can 
ignore the mixing-length parameter $\alpha$ which is relevant only
to the outer convective envelope and hence $R$ and \teff.
In Fig. \ref{fig:amlum} we show the evolution of luminosity
as a function of age, for models of different mass (denoted with a number)
and changing $Y_i$ and $Z_i/X_i$, with the reference model of 0.80 \msol\ having
$Y_i$ = 0.245 and $Z_i/X_i$ = 0.0005.
We also highlight the error bars associated with $L$ when we adopt
the reddened and unreddened bolometric flux (reddened is a higher value).
We have limited the age range to a minimum expected value, i.e. 
the minimal age of the old galactic clusters $\sim$10 Gyr.

Using only the luminosity, Fig.~\ref{fig:amlum} illustrates
the following: { 1)} 
%\begin{itemize
The mass of the star is between 0.78 and 0.84 \msol\
if we adopt a minimum $Y_i$ = 0.245, consistent with a primordial value
predicted by Big Bang nucleosynthesis.  {2)} 
Increasing $Z_i/X_i$ by 0.0002 (typical error, see next section) 
changes the age by about 0.3 Gyr.
{3)} The effect of increasing $Y_i$ to an upper limit of 0.260 
(= $Y_i + 3\Delta Y$, where $\Delta Y = 0.005$ is indicated in Fig.~\ref{fig:amlum}) leads to a corresponding decrease in the age of the 
reference model by $\sim$1.5 Gyr if everything else is left fixed,
and imposing an approximate age of the Universe would decrease the lower
limit in the mass to 0.75 \msol.
Figure~\ref{fig:amlum} depicts the approximate correlations between $M$ and 
$Y_i$, given just the luminosity constraint.
In this figure, we have denoted an upper age limit of $\sim 14$ Gyr 
(approximately the age
of the Universe) by the dotted horizontal bars for each $Y_i$ 
which imposes a lower limit
in mass.
We then depict the upper limit in mass when we assume that Halo stars should
have begun forming at approximately $>12$ Gyr (continuous lines),
and when we relax this constraint to the age of the youngest globular 
clusters of $>10$ Gyr (dashed lines).
Using this assumption along with the assumption that the 
initial helium abundance is low ($Y_i < 0.26$), we obtain  
a model-dependent mass of \hd\ of between 0.75 and 0.84 \msol.
This then implies \logg\ = 3.65 \pmm\ 0.03 by imposing $R$ from this work.
%Figure~\ref{fig:amlum} also shows that
%a precise measurement of the mass potentially allows a
%tight constraint on $Y_i$, and certainly can reduce the uncertainty on the age of the star.

\subsection{Refining the parameters using the angular diameter constraint \label{sec:test}}

\subsubsection{Constraining the mixing-length parameter $\alpha$\label{sec:alpha}}
The radius and the \teff\ of a star are extremely sensitive to the adopted
mixing-length parameter $\alpha$ which parametrizes 
convection in the convective envelope.
It is an adjustable parameter and it needs to be tuned to fit the observational
constraints.  
However, because of the correlations among different parameters and sets 
of input physics to models, 
in many cases it is left as a fixed parameter, and its value is set to the
value obtained by calibration of the solar parameters with the same physics. 
This is also the case for many isochrones that are available publicly.
While adopting $\alpha$ as the solar-calibrated value  
may be valid for stars that are similar to the Sun, for
metal-poor evolved stars (different \teff, \logg, and \feh) this is certainly not
the case as has been shown by  
%by \citet{cre12b} for the metal-poor objects \object{HD\,122563} 
%and \object{Gmb\,1830}, 
 {\citet{casa07}, for example,  who suggest a 
downward revision for metal-poor stars,}
or \citet{ker0861cyg} for \object{61\,Cyg}, or more recently 
by \citet{cre12b} for two metal-poor stars, and 
from asteroseismology for a few tens of stars, e.g. \citet{met1216cyg,mathur12,bonaca12}.
Fixing $\alpha$ with a solar calibrated value for \hd\ yields no models consistent
with observations.
However, as we know that the age of \hd\ is high ($>10$ Gyr) and its mass is relatively
well-determined as argued in the above paragraph ($0.75 < M < 0.84$ \msol), 
along with its metallicity, $\alpha$ is one of the free 
parameters that we can in fact constrain, as well as further 
constraining its age and mass.

In order to reproduce the $R$ and \teff\ of \hd, it was necessary to 
decrease the value of $\alpha$ to about 1.00 (solar value $\sim$2.0).  
Figure~\ref{fig:alpha} illustrates how the value of $\alpha$ is constrained
by the stellar evolution models.  
 {We show the HR diagram depicting the position of the error box of 
the reddened solution.  
Assuming \yi\ = 0.245 the mass is between 0.78 and 0.84 \msol\ (see Fig.~\ref{fig:amlum}, right panel).
We plot as continuous lines a 0.78 (blue), 0.80 (black), and 0.82 (green) 
\msol\ model with a mixing-length
parameter of $\alpha = 1.0$.
Increasing $\alpha$ to 1.30 results in the 0.78 \msol\ evolution track falling inside the 
observed constraints (blue dashed).  
Decreasing $\alpha$ to 0.70 for the 0.82 \msol\ also results in the track falling
within the error  box (green dashed).
Directly from the models we can establish a plausible range of $\alpha$ for 
a given mass for \yi\ = 0.245:\\
-- 0.78 \msol\ implies $1.10<\alpha < 1.50$,\\
-- 0.80 \msol\ implies $0.85<\alpha < 1.25$, and\\
-- 0.82 \msol\ implies $0.5 < \alpha < 0.9$.\\
The black dashed line above the error box shows a 0.84 \msol\ model with
a value of $\alpha = 0.5$.  
We were unable to make exolutionary tracks with values of $\alpha < 0.5$, thus
establishing a model-dependent lower limit on its value, and restricting
the mass of the model to $M \le 0.83$ \msol.}

 {We can summarize $M$ as a function of $\alpha$ by using the following
relation $M = 0.885 - 0.080\alpha$ (\pmm\ 0.01) \msol\  or alternatively as 
$\alpha = 11.013 - 12.449M$ (\pmm\ 0.2) where the errors are the 1$\sigma$
uncorrelated errors. 
With a mass between 0.78 and 0.83 \msol\ and $\alpha$ between 0.5 and 1.50,
we set our reference model to the mean mass 
$M = 0.805 \pm 0.01$ \msol\ which implies $\alpha = 1.00 \pm 0.20$.
This is the solution adopted for \yi\ = 0.245 (Table~\ref{tab:models1}), and 
for comparison with other solutions we 
adopt $\alpha = 1.00$ as the reference value.}

%Figure~\ref{fig:hr} shows the error boxes in the HR diagram 
%depicting the position of \hd\ while adopting zero-reddening (left) and 
%\av\ = 0.10 mag (right).
%We also highlight models that satisfy the observations (pass through the error box)
%to illustrate the central model parameters and their uncertainties.
%On the left panel we also show the error box when we increase the distance 
%by 1$\sigma$ (dotted boxes), while on the right panel the dotted box 
%illustrates the unreddened solution.

%The dotted 'R' lines show the limits in the radius  
%when it is calculated directly
%from $L$ and the spectroscopic \teff, given in Table~\ref{tab:radius}.

\subsubsection{Mass, initial helium, and age of the non-reddened star \label{sec:unred}}
Figure~\ref{fig:hr}, left panel shows the HR diagram error box and 
stellar models that satisfy or are close to the unreddened solution.
The most central model corresponds to a mass of $M = 0.78$ \msol, an initial
helium abundance of $Y_i=0.245$, initial metallicity $Z/X_i = 0.0005$, and $\alpha = 1.00$.
We also highlight models of other masses (blue) with the same 
$Y_i$, $Z/X_i$, and $\alpha$ that pass through or close to the error box.
Models considering different $\alpha$ and \yi\ (green dashed lines) are 
also shown. %in Fig.~\ref{fig:hr}.  %Keeping the same mass and changing each of the other par
 {We apply the approach from  
the previous section to establish a relationship between
$M$ and $\alpha$, and we also take \yi\ into account.  Given two of 
$M,\alpha,Y_i$ the third can be derived as }
%to 
%reproduce the central part of the error box.}
%and we obtain 
%$M = 0.843 - 0.067\alpha$ or $\alpha = 12.650 - 15.000M$, 
%where $0.78\le M\le 0.81$ (the $M = 0.82$ model requires $\alpha<0.5$).
%As \yi\ also has an effect on the age and position of the model we can 
%extend the previous equation to include a \yi\ dependence.}
%
%\end{document}
\begin{equation}
%M = 1.170 - 0.067\alpha -1.333 Y_i, 
M = 1.062 - 0.059\alpha -0.914 Y_i, 
\label{eqn:may1}
\end{equation}
\begin{equation}
%alpha = 17.550 - 15.000M - 20.000 Y_i, {\rm or}
\alpha = 18.053 - 17.000M - 15.500 Y_i, {\rm or}
\end{equation}
\begin{equation}
%_i = 0.878 - 0.750M - 0.050 \alpha 
Y_i = 1.152 - 1.082M - 0.063 \alpha 
,\end{equation}
%ameters also yields
%models that pass through the error box (green dashed).
where these equations must satisfy
$0.77 \le M \le 0.82$, $0.5 \le \alpha \le 1.5$ and $0.245 \le Y_i \le 0.260$.
Using the reference $\alpha = 1.0$ we give two solutions in 
Table~\ref{tab:models1} corresponding to \yi\ = 0.245  (model 1, M1), and \yi\ = 0.255 (model 2, M2), resulting in masses of 0.78 and 0.77 \msol.

To determine the age and other properties corresponding to 
these models we consider the properties of all of the models that
pass through the 1$\sigma$ error box.  
These properties are given in the lower part of Table~\ref{tab:models1}
along with their uncertainties.
For M1 we obtain an age of 13.68 Gyr with \mh\ = --2.03.

\begin{figure*}
\center{\includegraphics[width=0.48\textwidth]{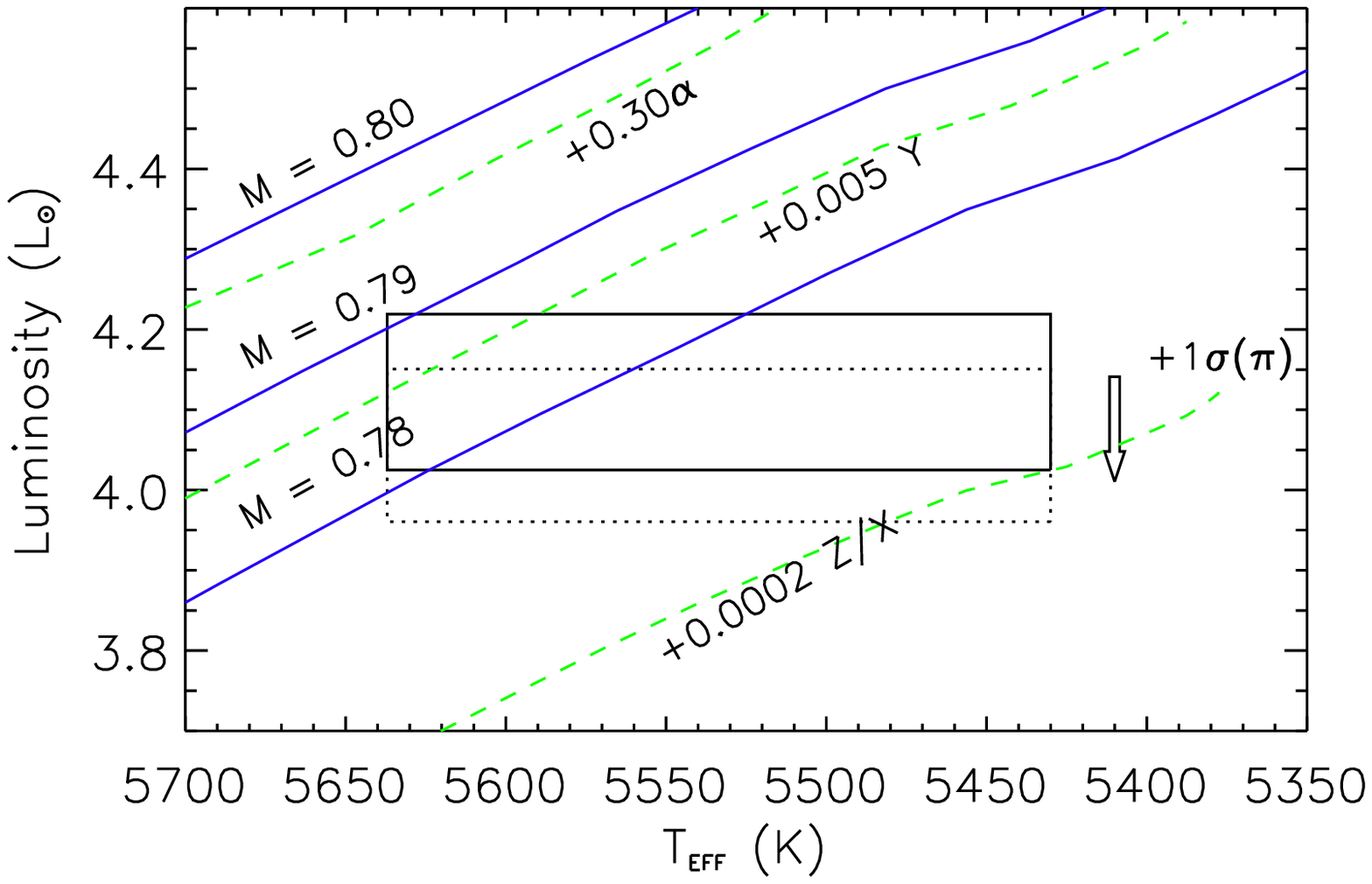}
\includegraphics[width=0.48\textwidth]{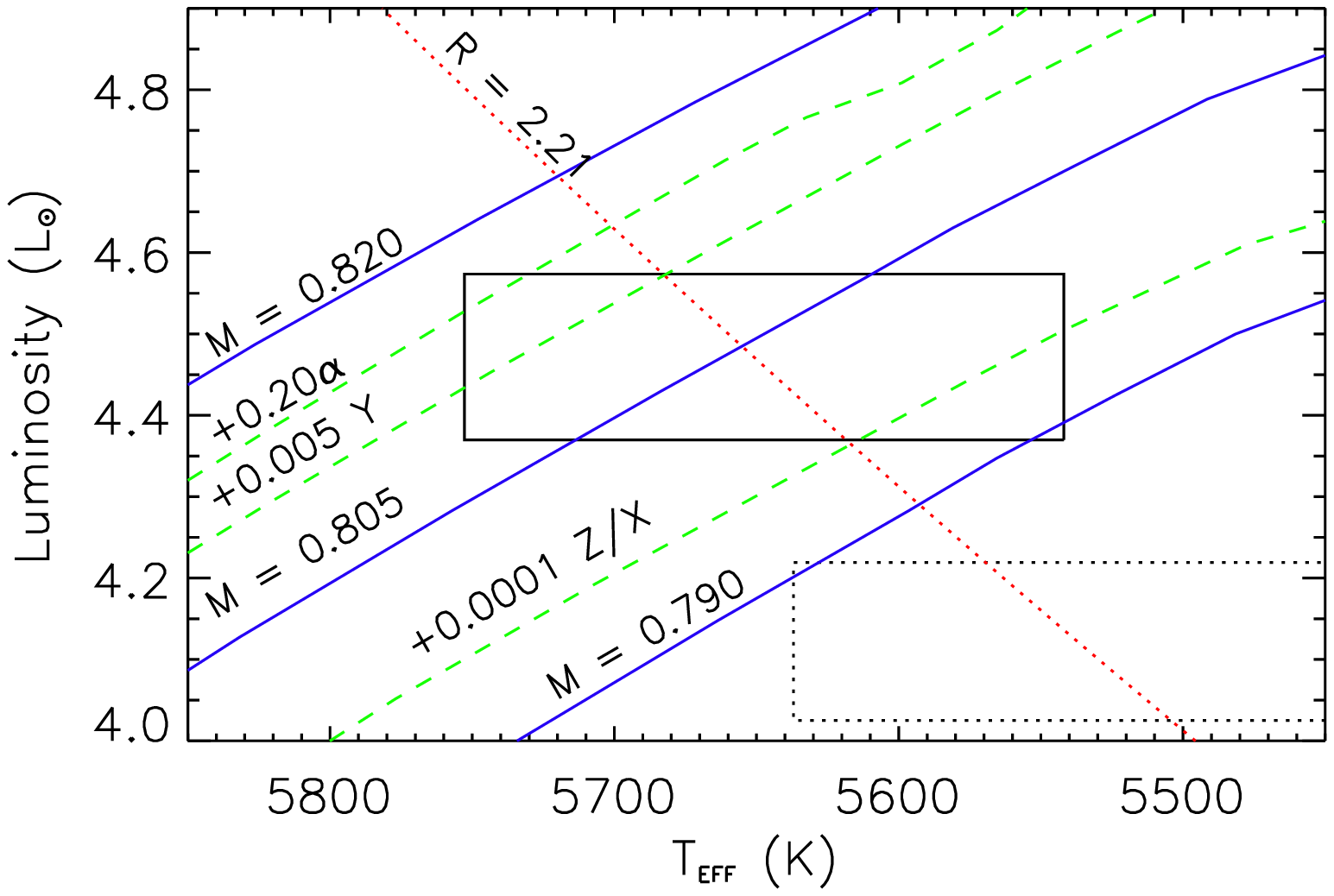}}
\caption{
HR diagrams illustrating the error box
corresponding to the observations derived in this work for \hd.  The left and right
panels show the results when we adopt \av\ = 0.00 and 0.10 mag.
The dotted box in the left panel shows the error box 
when the distance is
increased by 1$\sigma$, also illustrated by the arrow.
In the right panel (non-zero reddening solution) we also 
show the \av\ = 0.00 error box (dotted box),
located in the lower right part of the figure.  
The central models correspond to $Y_i= 0.245, \alpha = 1.00$, and $Z/X_i = 0.0005$ (with $Re_\nu = 5$).  
The optimal masses are $M= 0.780$ and $M=0.805$ \msol, and
are shown as the blue continuous lines passing near the centre of the error
boxes.
We also illustrate other models by changing the mass (blue) and the other parameters (green).
 The red dotted line corresponds to the radius derived in this paper.
 See Section~\ref{sec:test} for details.
 \label{fig:hr}}
\end{figure*}

\subsubsection{Mass, initial helium and age of the reddened star}
Figure~\ref{fig:hr} right panel illustrates the optimal models that satisfy the
$L$, \teff\ and \mh\ constraints (central continuous error box) for the 
reddened solution. 
The dotted error box shows the position of the unreddened constraints.
Following the approach above we 
establish equations for deriving one of $M, \alpha, Y_i$ as a function of
the other two.
\begin{equation}
% = 1.358 - 0.080\alpha - 1.933 Y_i
M = 1.121 - 0.080\alpha - 0.967 Y_i
\end{equation}
\begin{equation}
%alpha = 16.939 - 12.459 M - 24.153 Y_i
\alpha = 13.980 - 12.459 M - 12.077 Y_i
\end{equation}
\begin{equation}
%_i = 0.691 - 0.504 M -0.040\alpha
Y_i = 1.137 - 1.008 M -0.081\alpha
\label{eqn:may6}
\end{equation}
where the equations must satisfy
$0.78 \le M \le 0.83$, $0.5 \le \alpha \le 1.5$ and $0.245 \le Y_i \le 0.260$.
In Sect.~\ref{sec:alpha} we found an optimal model of $M = 0.805 \pm 0.010$ 
\msol\ for \yi\ = 0.245, and adopting the reference $\alpha = 1.00$, 
we obtain $M = 0.795$ \msol\ for \yi\ = 0.255.
The ages of these models are 12.17 and 11.95 Gyr respectively with
metallities of \mh\ = --2.06 and --2.07.
These properties are summarized in the lower part of Table~\ref{tab:models1}
under the column headings M3 and M4.
%We note that 

\begin{table}
\begin{center}\caption{Stellar model properties of \hd, adopting
$Y_i$ = 0.245 and 0.255, and $A_V = 0.00$ and 0.10 mag. 
See Sect.~\ref{sec:test} for details.
\label{tab:models1}
Ages and the corresponding $\alpha$ values for all combinations of model properties ($M$, $Y_i$, $A_V$) are given
in the on-line version.}
\begin{tabular}{lccccccccclllllllll}
\hline\hline
Parameter & Value & & Uncertainty\\% $\pm\sigma(M)$ & $\sigma(Z_i/X_i)$ & $\sigma(\alpha)$\\
$Y_i$ & 0.245 & 0.255 \\
\hline
\multicolumn{1}{l}{\av\ = 0.00 mag}& M1 & M2\\
\hline
$M$ (\msol) & 0.780 & 0.770 & 0.010\\
$Z_i/X_i$ & 0.0005 & 0.0005 & 0.0002\\
$\alpha$ & 1.00 & 1.00 & 0.20\\
$t$ (Gyr) & 13.68 & 13.45& 0.71\\
$\log g$ (cm s$^{-2}$)& 3.64 & 3.64& 0.03\\
%\meand\ (cm s$^{-3}$)&0.102&0.102&020\\
$L$ (\lsol) & 4.12 & 4.12& 0.10\\
\teff\ (K) & 5571 & 5601&  101\\
$R$ (\rsol) & 2.18 & 2.16& 0.08\\%$\langle \rho \rangle$ (g cm${-3}$) & 0.108 \pmm\ 0.015 & 0.110 \pmm\ 0.010\\
$[Z/X]_s$ (dex) & --2.03 &--2.05 &  0.07\\
\\
\hline
\multicolumn{1}{l}{\av\ = 0.10 mag}&M3&M4\\
\hline
$M$ (\msol) & 0.805 & 0.795 & 0.010\\
$Z_i/X_i$ & 0.0005 & 0.0005 & 0.0002\\
$\alpha$ & 1.00 & 1.00 & 0.20\\
$t$ (Gyr) & 12.17 & 11.95 & 0.62\\
$\log g$ (cm s$^{-2}$)& 3.65 &3.65 & 0.02\\
%\meand\ (cm s$^{-3}$)&0.102&0.103 & 0.014\\
$L$ (\lsol) & 4.42 & 4.42& 0.10\\
\teff\ (K) & 5662 & 5695&  105\\
$R$ (\rsol) & 2.20 & 2.18 & 0.08\\%$\langle \rho \rangle$ (g cm${-3}$) & 0.106 \pmm\ 0.017 & 0.105 \pmm 0.015\\
$[Z/X]_s$ (dex) & --2.06 & --2.07&  0.05\\
\hline\hline
\end{tabular}
\end{center}
\end{table}

\subsubsection{The global solution\label{sec:whichsolution}}
The determination of precise stellar properties that cannot be observed 
is extremely difficult, 
particularly because of the strong correlations between $M,\alpha$, and \yi, and 
consequently the age.
The luminosity constraint along with the stellar models and the assumptions on
the allowed age range of the star refined 
the mass to $0.75<M<0.84$ and consequently 
\logg\ to 3.65 \pmm\ 0.03.
The angular diameter then further constrained $\alpha$ to within a 
range of 0.5 -- 1.5 and $M$ to $0.77 < M < 0.83$, but most
importantly it provided a tight correlation between them, along with \yi.
Our models have a surface metallicity of \mh\ $\sim -2.05$, in good agreement with
the observed one (--2.10).

 {Determining the age of the star depends on the adopted values of 
of $M,\alpha,Y_i$, where only two are independent and the third is 
calculated from Eqs.~\ref{eqn:may1}--\ref{eqn:may6}.
The full range of mass values (by varying \yi\ and $\alpha$) 
has the largest impact on the determination of age, followed 
by \yi. The value of $\alpha$ only indirectly influences the age through the
required change of $M$ and \yi.
To attempt to quantify this effect, in Figure~\ref{fig:massagemodel} we show 
the ages of the models that
fit the observational constraints for a range of parameters.
We varied the mass (x-axis) and \yi\ (circles: \yi\ = 0.245, diamonds: \yi\ = 0.255) 
within the allowed correlations as shown in Fig.~\ref{fig:amlum}. 
For each combination we adjusted $\alpha$ to within its allowed range
until the evolution track
passed through the centre of the error box.
% while staying within the 
%allowed ranges, as specified in the paragraphs above.
This was done for the reddened solution (shown in red) and
the unreddened solution (black).
The figure shows the age of the models for these $M,Y_i$ combinations.
The crosses 
show the solutions in Table~\ref{tab:models1} (the central models
in Fig.~\ref{fig:hr}).}

 {Allowing for all plausible combinations of $M,\alpha,Y_i$ results in a possible
age of \hdd\ varying between 10.5 Gyr and 14.0 Gyr, with a lower limit of just under 11 Gyr 
for the zero reddened solution.
The required values of $\alpha$ are 
systematically lower for the zero reddened solution
with a typical value of $0.5 \le \alpha \le 1.00$ while the reddened solution
requires $0.7 \le \alpha \le 1.3$.}

We can parametrize the age $t$ as a function of $M,Y_i$, and \av\ using
the following equation
\begin{equation}
t = 79.8015-59.4171M-80.9789Y_i+0.7761A_v
\label{eqn:age}
,\end{equation}
where $0 \le A_V \le 0.10$, $0.245 \le Y_i \le 0.260$, $0.76 \le M \le 0.83$, 
and the combination of $M$ and \yi\ necessarily implies $\alpha$.
The value of $\alpha$ can be approximated by 
\begin{equation}
\alpha = 14.2038  -13.2170 M  -12.0251 Y_i+  3.7052A_v
\end{equation}
and is restricted to values between 0.5 and 1.5 (lower values 
for higher masses and vice versa).  The full range of ages and $\alpha$ for
all combinations of $M, Y_i, A_V$ are given in the on-line version.

Here it is clear that in order to determine the age of the star with the
best precision we need to restrict the values of 
one of the parameters $M,\alpha,Y_i$ (see Sect.~\ref{sec:discussion}).
Resolving the reddening problem 
would also relieve this degeneracy.

\begin{figure}
\includegraphics[width=0.48\textwidth]{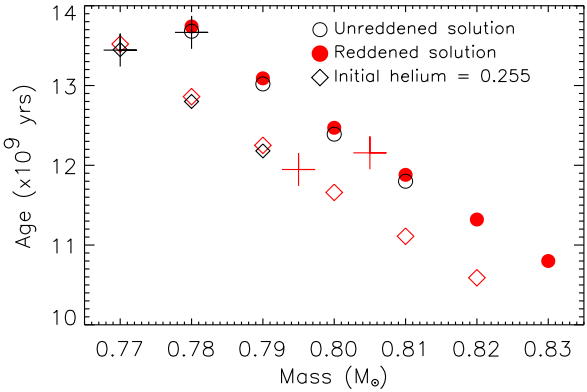}
\caption{\label{fig:massagemodel}The relationship between the age and the 
mass of the models that satisfies the observational constraints.  The circles/diamonds
correspond to the $Y_i = 0.245/0.255$ solution, and the red denotes the 
\av\ = 0.10 mag solution.  The crosses denote the values corresponding
to the centres of the error boxes in the HR diagram (Fig.~\ref{fig:hr})
and the solutions given in Table~\ref{tab:models1}. See Sect.~\ref{sec:whichsolution} for details.}
\end{figure}

 {Assuming that the mixing-length parameter does not vary much within the 
atmospheric parameters
of our models (\logg\ = 3.65, \mh\ $\sim -2.05$, \teff\ $\sim5530$ -- 5660 K), 
and by requiring that the different solutions adopt a similar value, we 
can judge the impact of reddening and \yi\ on the mass and age determination.
Table~\ref{tab:models1} shows four solutions that adopt $\alpha = 1.00$ for
two values of \yi\ and the two extreme values of \av.
The effect of increasing \yi\ by 0.01 is a decrease in the mass and the age by 0.01 \msol\
and $\sim 0.2$ Gyr.
The biggest difference comes from the adopted reddening, giving a 1.5 Gyr
difference between the solutions with \av\ = 0.0 and 0.1 mag.
In principle we may expect to be able to impose a value of $\alpha$ in the 
near future either from the calibration of this parameter with large samples of targets, e.g. 
from asteroseismology \citep{bonaca12,metcalfe14}, 
or from advances in 3D numerical techniques, e.g. 
\citet{tramp11}. Imposing this parameter externally would reduce the total 
range of mass, \yi\ , and possibly extinction, thus delivering a more reliable
determination of its age.}

The uncertainties that we give in Table~\ref{tab:models1} for $M$, \zx, and $\alpha$ 
are the 1$\sigma$ uncorrelated uncertainties.  These are obtained by fixing the parameters
and varying the one of interest until we reach the edge of the error box.  
The uncertainties in the other parameters are obtained by varying 
the above-mentioned parameters up to 1$\sigma$ at the same time and 
requiring that they remain within the error box, e.g. for solution M3 the mass 
took values of between (0.815 - 0.790)/2 \msol\ (rounded), while $\alpha$ varied between 
1.2 and 0.8, and \zx\ between 0.0003 and 0.0007.

\section{Discussion\label{sec:discussion}}

\subsection{Observational results}
In this work we determined three important quantities for \hd: the angular
diameter $\theta$, the bolometric flux \fbol, and a 1D LTE \teff\ derived from 
H$\alpha$ line wings.  Combining the first two observational
results with a well-measured parallax from \citet{bond13} yields the
stellar parameters $R$, $L$, \teff, and \logg.
In this work we found that the biggest contributor to our systematic error
in $L$ and \teff\ is the existence or not of 
interstellar reddening. The adoption of zero or non-zero reddening leads
to a change in $L$ and \teff\ by 
$\sim$0.3 \lsol\ and 100 K, respectively.
The determination of the spectroscopic \teff\ led to a value closer to the 
reddened solution (5626 K).
If we used this spectroscopic value to deduce \fbol\ 
and thus reddening (using the angular diameter),
we would obtain \fbol\ = 4.16 x 10$^{-8}$ erg s$^{-1}$ cm$^{-2}$, giving
a flux excess of $\sim$0.27 x10$^{-8}$ and corresponding to \av\ $\sim$ 0.084 mag or 
E(B-V) = 0.027 mag by adopting an extinction law of R(V) = 3.1.
This value would be in agreement with the value proposed by \citet{bond13}.

\subsection{1D LTE spectroscopic \teff}
In this work we determined a 1D LTE spectroscopic \teff\ using H$\alpha$ line wings
while imposing \logg\ = 3.65 \pmm\ 0.06 and \feh\ = --2.46 \pmm\ 0.14.  
The \logg\ was imposed from this work
using the interferometric \rad\ and an estimate of $M = 0.80$ \msol, later confirmed
with models to be correct.
The question we posed was whether such an analysis is capable of 
determining \teff\ of metal-poor stars where generally missing physics in atmospheric
models can lead to a large variety of \teff\ determinations with correlations between
the parameters.  
In Figure~\ref{fig:compareteff} we show determinations of the spectroscopic parameters from various
authors as provided by the PASTEL catalogue \citep{soubiran10}.  We show the [Fe/H]
determinations against \teff\ with a colour and symbol code corresponding 
to different \logg.  The blue filled circles are those with $3.6 < \log g < 3.7$ corresponding
to the constraints we can impose from this work (in many cases their \logg\ was constrained
prior to the analysis).  
The black circle marks our determination of \teff. % and \feh\ with \logg\ = 3.65.
%We can see that our spectroscopic \teff\ is consistent with the results whose \logg\ 
%constrained.
The shaded background illustrates the $\pm1\sigma$ determination of the interferometric 
\teff\ assuming reddening, and the dashed lines illustrate the boundaries with an assumption
of zero reddening.
If we can show that we need to consider reddening then we can conclude that the 
spectroscopic analysis is indeed capable of reproducing the \teff\ of such metal-poor stars,
if we can impose a reliable constraint such as \logg.

 {Recently \citet{ruchti13} studied systematic biases in \teff\ determinations,
%determined an independent \teff\ using a similar approach
and they found that the \teff\ determined from H$\alpha$ were 
hotter than those derived from the angular diameter
for metal-poor stars by 40 -- 50 K (including \hdd).
Our results are in agreement with
this statement if we assume some mild absorption to the star, 
e.g. $A_V \sim 0.05$ mag.   

For \hdd,\ \citet{ruchti13} 
% results was based on 3 stars, including \hdd\ where they
used surface brightness relations to estimate the 
interferometric \teff\ for this star
(5720 K), and they obtained an H$\alpha$ \teff\ = 5775 K, 150 K hotter than ours.
There are a few explanations for this difference: 1) the \logg\ and \feh\ values
adopted affect the derived \teff\ (see Table~\ref{tab:spectlogfeh}); 
however, they did not specify these values in their work; 2) the observations  
give different results --- our analysis with NARVAL data yields higher \teff\ than the other
three sets; 3) the theoretical profiles were calculated using different 
model atmospheres and code; and 4) self-broadening was computed using a
different (older) theory.  
We believe that our analysis using different line masks and four sets of observations
should yield a more reliable result.  
The only way to unveil the origin of the systematic difference is to 
use each other's observations and compare the results.}

%Alternatively if we have confidence in the spectroscopic determinations our results would point 
%towards a non-reddened solution and thus an optimal mass and age of $M=0.80$ \msol\ and 
%12.2 Gyr.

\begin{figure}
\includegraphics[width=0.45\textwidth]{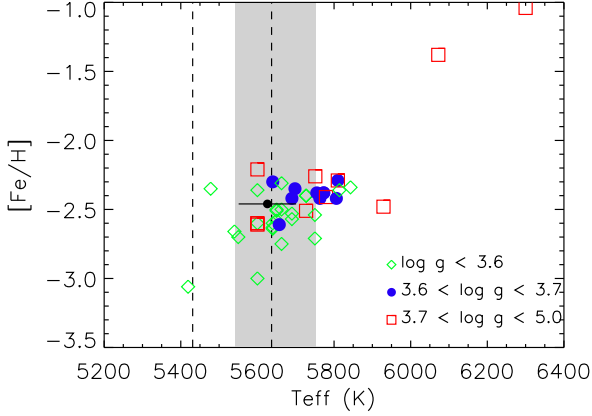}
\caption{Spectroscopic determinations of \teff\ and \feh\ from the literature (coloured symbols) 
and our
spectroscopic determination using H$\alpha$ line wings (black circle).
The shaded area illustrates the interferometric determination of \teff\ from the non-zero 
reddening solution, while the dashed lines illustrate the boundaries of the interferometric
\teff\ from the zero reddening solution.\label{fig:compareteff}}
\end{figure}

\subsection{Initial metallicity and the extra-mixing parameter in stellar models}

The mean surface metallicity of our most reliable models is $[Z/X]_s$ = --2.05, with an initial value of $[Z/X]_i$ = --1.70 or $Z_i/X_i$ = 0.0005 (the surface metallicity is mostly
set by the initial value with some slight variation due to $\alpha$).
This value was obtained by adopting a diffusion parameter in the stellar models, one that we determined by comparing the
difference in surface metallicity abundance of stars at their turn-off stage and the base of the giant
branch with observed results from \citet{nord12} of the globular cluster NGC~6397.
However, as it still remains  an adjustable parameter in the stellar
models, it could be incorrect.  Adopting 
a  much smaller value results in a maximum change in surface metallicity during main
sequence evolution of up to 1.0 dex (see e.g. Fig.~\ref{fig:diff} which
shows a maximum change of 0.55 dex).  This would require an initial
value of the order of --1.40 dex.  
This scenario is improbable and would result in a surface metallicity difference
between turn-off and giant stars 
in disagreement
with \citet{nord12}. 
Adopting a larger value for the parameter would reduce the maximum change 
in surface metallicity
during evolution, 
and converging on a surface value of --2.10 dex would require  
the initial value to be decreased, for example to --1.90 dex or $Z_i/X_i$ = 0.0002.
For a star with an age almost the age of the Galaxy (adopting zero reddening) 
a lower initial metallicity could be more likely. 
However, a star 1 billion years younger could be consistent
with a higher initial metallicity.
The impact of this change of physics and 
metallicity on the resulting stellar properties is currently limited
by our observational errors.
At the moment we cannot explore this  further and we are required to impose
the external constraints.

\subsection{Improving the precision and accuracy of the age}
It is clear that in order to determine the most accurate and precise age for 
\hd, it is necessary to solve the extinction problem, and to 
reduce the span of masses that pass through the error box.
The second question can be addressed 
by obtaining 
a more precise determination of the angular diameter. % should be determined.
 From Fig.~\ref{fig:hr} one can immediately see that by reducing the
error bar in \teff\ by a factor of two, fewer models would satisfy the constraints.
As an example,  the right panel shows that models of masses between 0.790 and 
0.805 (a total span of 0.015 \msol) would be the only models that 
pass through the error box of half of its size.  
The parallax and \fbol\ already contribute very little to the error bar
for \rad\ and \teff, and so the only option is to obtain a precision 
in \thet\ of the order of 0.007 mas. 
By cutting both the external and the statistical errors in half, this 
precision could be achieved. %The error on the angular diameter would 
The current precision is just under 4\%, which is excellent 
considering that we are already working at the limits of angular 
resolution using the CHARA array with VEGA and VEGA's sensitivity limits.
However, the telescopes of the CHARA array will be equipped with adaptative optic (AO) systems 
in the next one or two years. This will improve both the sensitivity and
measurement precision of VEGA. 
The gain in sensitivity would allow us to observe fainter 
(so smaller) calibrators, hence reduce the external error 
affecting the
        calibrated visibilities. 
New observations of HD140283 with CHARA/VEGA or CHARA/FRIEND (future instrument)
equipped with AO would allow us to achieve the necessary precision.

\begin{figure}
\includegraphics[width=0.48\textwidth]{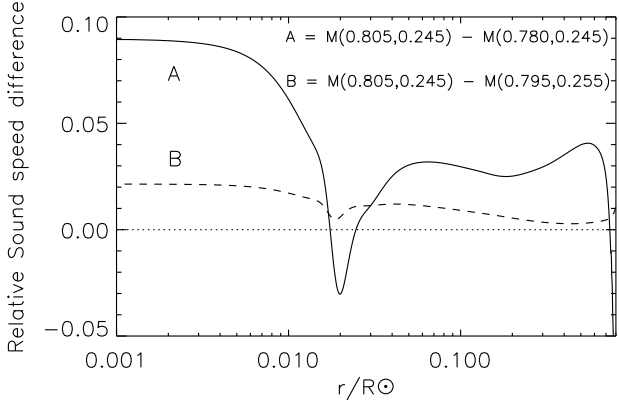}
\caption{\label{fig:sspeed} Relative difference in the sound speed profile between models M3 and M1 (continuous) and M3 and M4 (dashed). }
\end{figure}

\subsection{Improving the mass and age from asteroseismology}
Another way to determine the mass of the star is through the detection
and interpretation of stellar oscillation frequencies.  
Acoustic oscillation frequencies (sound waves) are sensitive to the 
sound speed profile of the star and thus its density structure and mean density.
Because \rad\ has been measured, 
%then $M$ can be determined with
%high accuracy through model-fitting
%given that the density is very well constrained.
%estimated independently
%of models using the so-called {\it scaling relations}.
in-depth asteroseismic analysis would allow a very high precision model-dependent 
determination of the mass.  
In Fig.~\ref{fig:sspeed} we show the relative differences between the 
sound speed profiles (from 0.02 to 0.8 the stellar radius) for the M3 and M1 models, 
(M3-M1)/M3.  This represents the difference between the reddened and unreddened solution
for $Y_i = 0.245$ (continuous).
At the same time, we also show the difference between the reddened solutions with different initial
helium abundance, M3 and M4 (dashed).  
The largest differences for both cases 
are found very close to the centre of the star
and near the outer convection zone.
These differences cause changes in the frequencies of the oscillation modes
that can  be detected, e.g. a radial high-order mode ($l=0,n=24$) has a predicted
frequency of 1240.2 $\mu$Hz for M3 and this same mode for M1 is 1251.0 \mhz,
although for the M4 model (different $Y_i$) it is 1240.7 \mhz, 
not distinguishable
from the M3 solution at current asteroseismic modelling 
limitations (e.g. \citealt{kje08}).
Precision in asteroseismic frequencies are well below 1.0 \mhz, even
for a ten-day ground-based observational campaign.
Stellar oscillations have been detected in stars of this mass.
At this stage of evolution, the so-called g-modes are very 
sensitive to the core conditions and hence should be able to distinguish even 
more clearly between both models.
 {However, we need to verify whether there are sufficient metallic lines in the spectrum to 
accurately detect radial velocity shifts (the oscillations).}
% and if a star
%with such low metallicity is capable of sustaining oscillations.

\subsubsection{Detection of oscillations in metal-poor stars}
The star $\nu$ Ind has a similar mass to HD\,140283, but 
has a higher metallicity ([M/H] = --1.3).  This star was the subject of 
a radial velocity asteroseismic campaign using telescopes in Australia and Chile 
and solar-like oscillations were detected
\citep{bed06,car07}.  
The oscillations were analysed and the amplitudes of the frequencies
were in the range 50-170 cm s$^{-1}$ (the solar amplitudes are of the order of 23 cm s$^{-1}$,
see also \citet{kje08b} for a more recent analysis of the amplitudes of both of these
stars).
A recent asteroseismic observational campaign of a different 
star using the SOPHIE spectrograph at the 
Observatoire Haute Provence in February 2013 allowed us to 
reach a precision of 20-25 cm~s$^{-1}$ in the power spectrum for a total of 8 good nights out
of 12 assigned.  
The detections of such oscillation frequencies are clearly very feasible using 
different instruments or telescopes.
Oscillations are detectable in metal-poor stars, but how metal-poor is observationally 
possible?
\hd\ would be an ideal target for continued observations,  {because 
the non-detection of oscillations would impose an upper limit on the 
amplitudes of the modes in metal-poor stars,}
%on the metallicity of the star capable of sustaining oscillations,
while the detection of oscillation frequencies would provide valuable 
constraints on the mass and age of this star, and stellar interior models.

\section{Conclusions}

In this work we determined stellar properties of \hd\ by 
measuring the angular diameter of the star and performing bolometric flux fitting.
The measured diameter is $\theta_{\rm 1D} = 0.353 \pm 0.013$ mas and coupling this
with a parallax of $\pi = 17.15 \pm 0.14$ mas from \citet{bond13} 
we obtain a linear 
radius of the star $R = 2.21 \pm 0.08$ \rsol\ and \logg\ = 3.65 \pmm\ 0.06.
The bolometric flux fitting resulted in two solutions, one where interstellar
extinction \av\ = 0.0 mag, and the other with a maximum non-zero value of 0.1 mag.
Adopting these two values we derived $L$ (4.12 or 4.47 \lsol), \teff\
(5534 or 5647 K), bolometric corrections,
and the absolute magnitude $M_V$ (Table~\ref{tab:radius}). 
We also determined a spectroscopic \teff\ using a 1D LTE analysis fitting
H$\alpha$ line wings and we found a value of {5626 K},
%a value more compatible with \teff\ derived using \av\ = 0.10 mag (5647 K).
a value more compatible with the interferometric \teff\ assuming a small amount
of reddening.
If \fbol\ is indeed a higher value because of interstellar reddening,
this is a very important result as it shows that a 1D LTE analysis
of H$\alpha$ wings yields accurate \teff\ at this low-metallicity regime.
%Adopting the spectroscopic \teff\ and $L$ we determined a spectroscopic
%$R$ of 2.15 and 2.24 \rsol, and a \logg\ of 3.68 and 
%3.64 dex (see Table~\ref{tab:radius}).
Our measured angular diameter is larger by about 2$\sigma$ than that predicted by
the surface brightness relations of \citet{kervella04b}, suggesting that indeed more
determinations of \thet\ are needed for such metal-poor and/or evolved objects 
to more accurately calibrate this scale,  especially important with the advent of 
Gaia and with the availability of precise distance measurements   for a billion
stars in the next five years.

 {By interpreting the observational results using stellar models we 
derived a strict relationship between mass $M$, mixing-length parameter $\alpha$,
 and initial helium abundance \yi\ (Eqs.~\ref{eqn:may1}--\ref{eqn:may6}),
where only two are independent.
We further determined the age $t$ of the star for all 
possible combinations of these parameters; Eq.~\ref{eqn:age} 
parametrizes $t$ as a function of the adopted $M$, \yi\ (implying 
$\alpha$), and reddening \av.
Fixing $\alpha = 1.0$ we studied the impact of changing \yi\ and 
\av\ on the final solution. 
For \av\ = 0.10 mag we obtain 
$M$ = 0.805 \pmm\ 0.010 \msol, corresponding
to an age of 12.17 \pmm\ 0.69 Gyr. % (Table~\ref{tab:models1}).
For \av\ = 0.00 mag we obtain a slightly lower mass 
of $M$ = 0.780 \pmm\ 0.010 \msol, corresponding 
to an age of 13.68 \pmm\ 0.61 Gyr (fixing \yi\ = 0.245). %, with $L$ = 4.18 \lsol\ and \teff\ = 5546 K.
The impact of \yi\ is much smaller, causing a decrease in $M$ and $t$ of 
0.01 \msol\ and 0.2 Gyr.
Table~\ref{tab:models1} summarizes some of our modelling results, while the
on-line version gives the full range of parameter solutions.

By adopting
zero or very little reddening \citet{bond13} and \citet{vandenberg14} 
derive an age of $\sim$14.3 Gyr in a completely independent manner.
These results were obtained by adopting an oxygen abundance
which was derived with a higher \teff\ than the one determined in this work.
In their work, the mixing-length parameter was not explored in 
their models and this would shift the evolution tracks towards lower \teff,
thus providing a match to the observational data at a lower age.
%, although with observational
%constraints.
\citet{casagrande2011} in an independent manner 
derive an age of 13.8 Gyr and 13.9 Gyr assuming
no reddening and using the BaSTI \citep{basti07} and PADOVA \citep{padova08} 
isochrones.
These determinations of the age of \hdd\ give an 
indication of the systematic error we can 
expect on stellar ages by using different approaches and assumptions on
both the data and models.

For all of our stellar models we needed to adapt the mixing-length parameter
to a value much lower than the Sun, i.e. from 2.0 to 1.0 using the \citet{egg72}
treatment of mixing-length.   If reddening is non-existent, then
$\alpha$ takes   a lower value ($0.5 \le \alpha \le 1.0$) on average. 
This result reinforces previously published 
results of the non-universality of the mixing-length parameter, e.g. 
\citet{miglio05}, one that 
is generally fixed in published stellar evolution tracks/isochrones.
 {We note that our somewhat lower \teff\ (compared to some recent determinations
of 5700 K and higher)
along with the observed Li abundances provides an important
reference point for studying Li versus \teff\ for metal-poor stars 
(see e.g. \citet{char05} , their Fig.~12) and consequently Li surface depletion
scenarios.}
%where such high abundances (A(Li) = 2.15 --- 2.28) seem to be more
%characteristic of higher \teff\ stars, bringing into question the 
%initial Li abundances and Li surface depletion scenarios.

The next crucial steps to better understand this star and to decorrelate
the stellar properties of mass, age, initial helium abundance, and  to 
look at the treatment of diffusion in such stars, is to reduce the 
uncertainty on the angular diameter by at least a factor of two, and to 
perform asteroseismic observations to better recover the internal
profile of the star and hence its mass and age.
 {Predictions of the mixing-length parameter from external constraints would 
also lead to a large improvement in determining mass, initial helium, and age.
All of these are realistic with current numerical capabilities 
and current or near-future instrumentation. }
One of the biggest improvements to the age and mass determinations, however, relies 
on the exact knowledge of extinction between the Earth and this star.

%__________________________________________________________________

%\section{Fundamental parameters and stellar modelling \label{sec:fparams}}

%__________________________________________________________________

\begin{acknowledgements}

This paper is dedicated to the wonderful astronomer Olivier Chesneau.
We thank Isabelle Tallon-Bosc for detailed discussions regarding 
uncertainties in interferometric data.
We thank the VEGA observing group at the 
Observatoire de la C\^ote d'Azur (OCA) in Nice, 
France 
(remote observing) and at CHARA, California, USA.
The work by the 
SAM (\url{http://www.astro.uu.se/\~ulrike/GaiaSAM/}) group in defining a set
of benchmark stars is greatly appreciated.
The CHARA Array is funded by
the National Science Foundation through NSF grant AST-0908253 and 
by Georgia State University through
the College of Arts and Sciences. 
TSB acknowledges support provided through NASA grant ADAP12-0172.
UH acknowledges support from the Swedish National Space Board (Rymdstyrelsen).
PK acknowledges the support of PHASE, the high angular resolution 
partnership between
ONERA, Observatoire de Paris, CNRS and the Universit\'e Denis Diderot, Paris 7.
We acknowledge financial support from the ``Programme National 
de Physique Stellaire" (PNPS) of CNRS/INSU, France.
During part of this research OLC was a Henri Poincar\'e Fellow at OCA,
which is funded by the 
Conseil G\'en\'eral des Alpes-Maritimes and OCA.
\end{acknowledgements}

\bibliographystyle{aa}
\bibliography{metal}

\appendix
\section{Bolometric flux fitting approach 2 (method 2B and 2C)\label{sec:orlaghfbol}}
 We describe in detail the second bolometric flux-fitting
approach that was implemented in this work.
The method  
% {An independent fitting method (2) was also used to determine \fbol\ and \av.
incorporates a non-linear least-squares minimization algorithm (Levenberg-Marqwardt) 
to find the best scaled interpolated spectrum that fits a set of  
observed flux points or photometric magnitudes (in this work we just use the flux data).
The fitting method requires on input a set of 
initial parameters --- \teff, \feh, \logg,
E(B-V), and a scaling factor $\theta_{\rm s}$ --- 
which define the characteristics of 
the spectra, the reddening to apply, and ratio of the stellar radius to 
the star's distance.  These parameters are optimized within
the algorithm and the bolometric flux is calculated by numerical integration 
of the scaled optimal spectrum.
The method is designed to allow implementation of any spectral library.
In this work we use the BASEL \citep{lej97} and PHOENIX libraries
\citep{phoenix-orig,phnx2013}.

Because of the strong degeneracy between the stellar parameters,
in practice we fix three of the values \teff, \logg, \feh, \av, while 
fitting the fourth along with $\theta_{s}$, so only two stellar parameters
are fitted at any one time.
Changing \logg\ and \feh\ by 0.5 dex, which is much larger than the typical 
errors on these values, has a negligible influence on the final 
calculated bolometric flux \fbol.  Generally these parameters are 
the ones that are fixed. 
Changing \av, however, does have an important effect as explained below.

{Within the fitting process, each spectrum is linearly interpolated 
between adjacent parameter points to the desired parameter
values as proposed by the minimization algorithm.
We have tested for different interpolation schemes and because of the 
relatively dense grid spacing (e.g. 100 K or 250 K in \teff, 0.5 in \logg,
0.5 in \feh) the choice and order of interpolation is irrelevant.
The scale factor $\theta_{\rm s}$ is also proposed at the same time
by the minimization algorithm.  
A \chisqr\ value is then calculated between the observed data
and the newly proposed scaled spectrum.
If the \chisqr\ value improves then the minimization algorithm
continues in the same direction.  
When the \chisqr\ fails to improve after four iterations the fitting stops,
but convergence is usually reached within only three or four iterations.
The final results are not sensitive to the initial parameters; 
for example, 
initial \teff\ values differing by 1000 K yielded consistent results.

To account for reddening we apply reddening laws to the spectra 
using the Interactive Data Language (IDL) routine {\tt ccm\_unred} and the input E(B-V), along with
a choice of extinction laws of $R_V$, which we fix at 3.1 in this work. 
This routine is based on the \citet{ccm89} coefficients and 
updated with \citet{odonnell94} in the UV. The values of
\teff\ and \av\ are highly degenerate parameters if no 
external constraints are imposed, and so 
to determine \av\ a fitting process is begun allowing only
\teff\ and $\theta_{\rm s}$ to be fitted, but imposing
a range of values of E(B-V). 
The fit yielding the smallest
\chisqr\ defines the adopted E(B-V) and thus \av\ (using $R_V$).  
The same results are obtained by fixing \teff\ at different 
values while fitting E(B-V).
If the observed data give a best match to a reddened 
spectrum, the intrinsic (unreddened) flux of the star is thus larger,
indicating a more luminous star.

\end{document}